\documentclass{IEEEtran}
\usepackage{times,amssymb,amsmath,amsfonts,float,nicefrac,color,subfigure,algorithm2e}
\usepackage{euscript,graphics}
\usepackage[dvips]{epsfig}
\usepackage{graphicx}
\newtheorem{theorem}{Theorem}[section]
\usepackage[]{algorithm2e}
\newtheorem{lemma}[theorem]{Lemma}
\newtheorem{proposition}[theorem]{Proposition}
\newtheorem{corollary}[theorem]{Corollary}

\newtheorem{cnstr}{Construction}

\newtheorem{xmpl}{Example}

\newcommand{\remove}[1]{}
\renewcommand{\tilde}{\widetilde}

\newcommand\nd{\noindent}

\newcommand\wt{\mbox{{\rm wt}}}

\def\rank{\qopname\relax{no}{rk}}

\newcommand{\ceilenv}[1]{\left\lceil #1 \right\rceil}

\newcommand\nc\newcommand
\nc\bfa{{\boldsymbol a}}\nc\bfA{{\bf A}}\nc\cA{{\mathcal A}}
\nc\bfb{{\boldsymbol b}}\nc\bfB{{\bf B}}\nc\cB{{\mathcal B}}
\nc\bfc{{\boldsymbol c}}\nc\bfC{{\bf C}}\nc\cC{{\mathcal C}}
\nc\bfd{{\boldsymbol d}}\nc\bfD{{\bf D}}\nc\cD{{\mathcal D}}
\nc\bfe{{\boldsymbol e}}\nc\bfE{{\bf E}}\nc\cE{{\mathcal E}}
\nc\bff{{\boldsymbol f}}\nc\bfF{{\bf F}}\nc\cF{{\mathcal F}}
\nc\bfg{{\boldsymbol g}}\nc\bfG{{\bf G}}\nc\cG{{\mathcal G}}
\nc\bfh{{\boldsymbol h}}\nc\bfH{{\bf H}}\nc\cH{{\mathcal H}}
\nc\bfi{{\boldsymbol i}}\nc\bfI{{\bf I}}\nc\cI{{\mathcal I}}
\nc\bfj{{\boldsymbol j}}\nc\bfJ{{\bf J}}\nc\cJ{{\mathcal J}}
\nc\bfk{{\boldsymbol k}}\nc\bfK{{\bf K}}\nc\cK{{\mathcal K}}
\nc\bfl{{\boldsymbol l}}\nc\bfL{{\bf L}}\nc\cL{{\mathcal L}}
\nc\bfm{{\boldsymbol m}}\nc\bfM{{\bf M}}\nc\cM{{\mathcal M}}
\nc\bfn{{\boldsymbol n}}\nc\bfN{{\bf N}}\nc\cN{{\mathcal N}}
\nc\bfo{{\boldsymbol o}}\nc\bfO{{\bf O}}\nc\cO{{\mathcal O}}
\nc\bfp{{\boldsymbol p}}\nc\bfP{{\bf P}}\nc\cP{{\mathcal P}}
\nc\bfq{{\boldsymbol q}}\nc\bfQ{{\bf Q}}\nc\cQ{{\mathcal Q}}
\nc\bfr{{\boldsymbol r}}\nc\bfR{{\bf R}}\nc\cR{{\mathcal R}}
\nc\bfs{{\boldsymbol s}}\nc\bfS{{\bf S}}\nc\cS{{\mathcal S}}
\nc\bft{{\boldsymbol t}}\nc\bfT{{\bf T}}\nc\cT{{\mathcal T}}
\nc\bfu{{\boldsymbol u}}\nc\bfU{{\bf U}}\nc\cU{{\mathcal U}}
\nc\bfv{{\boldsymbol v}}\nc\bfV{{\bf V}}\nc\cV{{\mathcal V}}
\nc\bfw{{\boldsymbol w}}\nc\bfW{{\bf W}}\nc\cW{{\mathcal W}}
\nc\bfx{{\boldsymbol x}}\nc\bfX{{\bf X}}\nc\cX{{\mathcal X}}
\nc\bfy{{\boldsymbol y}}\nc\bfY{{\bf Y}}\nc\cY{{\mathcal Y}}
\nc\bfz{{\boldsymbol z}}\nc\bfZ{{\bf Z}}\nc\cZ{{\mathcal Z}}
\nc\od{{\bar d}}\nc\ow{{\bar w}}\nc\odelta{{\bar\delta}}
\nc\ox{{\bar x}}\nc\oy{{\bar y}}\nc\ou{{\bar u}}
\nc\oh{{\bar h}}
\DeclareMathOperator{\E}{\text{\sf E}}

\DeclareMathOperator{\Cl}{Cl}

\newcommand\ff{{\mathbb F}}

\nc\ellone{{\ell_1}}
\nc\elltwo{{\ell_2}}
\nc\ellinf{{{\ell_\infty}}}
\nc\ip[2]{\langle #1,#2\rangle}

\newcommand{\beeq}{\begin{eqnarray*}}
\newcommand{\eneq}{\end{eqnarray*}}

\newcommand{\half}{\nicefrac12}

\begin{document}
\nc\red[1]{\textcolor{red}{#1}}

\title{Bounds on the Parameters of Locally Recoverable Codes} 
\author{Itzhak Tamo\IEEEauthorrefmark{1}\thanks{\IEEEauthorrefmark{1}I. Tamo is with the Dept. of EE-Systems, Tel Aviv University, Tel Aviv, Israel. The research was done while at the Institute for Systems Research, University of Maryland, College Park, MD 20742 (email: tamo@post.tau.ac.il). 
Research  supported in part by NSF grant CCF1217894.}
\hspace{0.7cm} \and Alexander Barg\IEEEauthorrefmark{2}\thanks{\IEEEauthorrefmark{2}A. Barg is with the Dept. of ECE and ISR, University of Maryland, College Park, MD 20742 and IITP, Russian Academy of Sciences, Moscow, Russia (email: abarg@umd.edu). Research  supported by NSF grants CCF1422955, CCF1217894, and CCF1217245.}
\hspace{0.7cm}\and Alexey Frolov\IEEEauthorrefmark{3}\thanks{\IEEEauthorrefmark{3}A. Frolov is with IITP, Russian Academy of Sciences, Moscow, Russia (email alexey.frolov@iitp.ru). Research supported by the Russian Science Foundation (project no. 14-50-00150). Part of this research done while visiting Institute for Systems Research, University of Maryland, College Park, MD 20742.}
\thanks{A part of the results of this paper were presented at the 2014 IEEE International Symposium on Information Theory, Honululu, HI \cite{Tamo14}.}
}

\maketitle
\remove{\author{\IEEEauthorblockN{Itzhak Tamo$^a$}
\and \IEEEauthorblockN{Alexander Barg$^{b}$}
\and \IEEEauthorblockN{Alexey Frolov$^{c}$}
}

{\renewcommand{\thefootnote}{}\footnotetext{

\vspace{-.2in}
 
\noindent\rule{1.5in}{.4pt}

\nd$^\ast$ Department of ECE and Institute for Systems Research, University of Maryland, College Park, MD 20742. Emails: 
\{zactamo,alexanderbarg\}@gmail.com. Research supported by NSF grants CCF1217894, CCF1217245, and by NSA grant 98230-12-1-0260.
}}
\renewcommand{\thefootnote}{\arabic{footnote}}
\setcounter{footnote}{0}

}

\begin{abstract}
A locally recoverable code (LRC code) is a code  over a finite alphabet such that every symbol in the encoding is 
a function of a small number of other symbols that form a recovering set. In this paper we derive new finite-length and 
asymptotic bounds on the parameters of LRC codes. For LRC codes with a single recovering set for every coordinate, 
we derive an asymptotic Gilbert-Varshamov type bound for LRC codes and find the maximum attainable relative distance 
of asymptotically good LRC codes. Similar results are established for LRC codes with two disjoint recovering sets for
every coordinate. 
For the case of multiple recovering sets (the availability problem) we derive a lower bound on the parameters using expander graph arguments.
Finally, we also derive finite-length upper bounds on the rate and distance of LRC codes with multiple recovering sets.
\end{abstract}

\section{Introduction}
Locally recoverable (LRC) codes currently form one of the rapidly developing topics in coding theory because of their applications in
distributed and cloud storage systems. Recently LRC codes have been the subject of numerous publications, among them
\cite{Gop11,Pap12,sil13,Tamo13,Gop13,matroid-paper,rawat2012optimal,Lluis2013}. 
Let $Q$ be a $q$-ary alphabet.
We say that a code $\cC\subset Q^n$ has locality $r$ if every symbol of the codeword $x\in \cC$ 
can be recovered from a subset of $r$ other symbols of $x$ (i.e., is a function of some other $r$ symbols $x_{i_1},x_{i_2},\dots,x_{i_r}$)  \cite{Gop11}.
In other words, this means that, given $x\in \cC, i\in [n],$ there exists a subset of coordinates 
${\cR}_i\subset [n]\backslash i, |{\cR}_i|\le r$ such that
the restriction of $\cC$ to the coordinates in ${\cR}_i$ enables one to find the value of $x_i.$ The subset 
${\cR}_i$ is called a {\em recovering set} for the symbol $x_i$. 
Generalizing this concept, assume that every symbol of the code $\cC$ can be recovered from $t$ disjoint subsets of symbols of size $r_1,\dots,r_t$ respectively.
Below we restrict ourselves to the case $r_1=\dots=r_t=r$ which makes the bounds obtained in the paper more compact. At the same time, 
we note that the technique presented below enables us to treat the general case as well.

Given a code $\cC\subseteq Q^n$ of size $q^k$ with $t$ disjoint recovering sets of size $r$, we use the notation 
$(n,k,r,t)$ to refer to its parameters. If the values of $n,k,r$ are understood, we simply call $\cC$ a $t$-LRC code.

More formally, denote by $\cC_I$ the restriction of the code $\cC$ to a subset of coordinates $I\subset [n].$
Given $a\in Q$ define the set of codewords
   $
   \cC(i,a)=\{x\in \cC: x_i=a\},\; i\in[n].
   $

\vspace*{.05in}
\nd{\sc Definition:} A code $\cC$  is said to have $t$ disjoint recovering sets if for every $i\in[n]$ there are $t$ pairwise disjoint subsets ${\cR}_{i}^1,\dots,{\cR}_{i}^t\subset [n]\backslash i$ such that for all $j=1,\dots,t$ and every pair of symbols $a,a'\in Q, a\ne a'$
  \begin{equation}\label{eq:rc}
  \cC(i,a)_{{\cR}_{i}^j}\cap \cC(i,a')_{{\cR}_{i}^j}=\emptyset.
  \end{equation}
 
 Having more than one recovering set is beneficial in practice because it enables more users to access a given portion of data, thus enhancing data availability in the system. 
 

In this paper we study upper and lower bounds on the parameters of $t$-LRC codes. Most of
our results concern bounds on the attainable value of the minimum distance $d$ of a code $\cC$ given its parameters
$(n,k,r,t).$
Since the main goal of LRC codes is to recover from one erased coordinate using its recovering set, it is 
not clear why one is interested in large values of the minimum distance. It is possible that
more than one storage nodes have failed, necessitating higher separation of the codewords, but the probability
of this event under the normal functioning of the system is low. To justify this problem from the 
perspective of applications, consider the situation when a cluster of nodes becomes inoperable due to either power failure or
maintenance. In this case it is desirable to be able to switch from local to global decoding, and this is
where large distance of the code becomes a useful feature.

A note on terminology: when we speak of lower bounds, our goal is to show that there exist codes, or sequences of
codes, that attain a particular relation between the parameters (e.g., have large distance). In the case of upper
bounds we aim to show that no code with given locality properties can have distance or rate greater than some
function of the other parameters of the code. In the proof of the upper bounds we do not make any assumptions on
the alphabet $Q$, while the lower bounds are proved using linear codes over finite fields.

We note that the case of $t=1$ is by far the easiest because good LRC codes with high distance are well structured. 
However even in this case lower bounds were largely absent from the literature. Namely, in the classic case, the asymptotic bounds for error-correcting codes pinpoint the value of the relative distance $\delta_0=(q-1)/q$ such that there
exist asymptotically good codes for all smaller $\delta$, and there are no code sequences with positive rate
for  $\delta\ge\delta_0.$ 
 In this paper we remedy this situation by deriving a Gilbert-Varshamov (GV) type bound that implies the same conclusion
for any constant value of $r$ (concurrent with our work, this question was also resolved in \cite{Cad13a}).

For codes with multiple recovering sets deriving bounds on the parameters is more involved because of the
mutual interaction between the sets that is difficult to quantify. For this problem we obtain the following results. 
First, we derive an upper bound on the maximum attainable rate of a $t$-LRC code expressed in terms of $r$ and $t$. We also derive an upper bound on the minimum distance of $t$-LRC codes given the
cardinality of the code and the value of the locality parameter. 
Turning to lower bounds, we derive 
an asymptotic GV-type bound on the parameters of codes with $t=2$ disjoint recovering sets. This result
again enables us to conclude that asymptotically good binary $2$-LRC codes exist only if the relative
distance $\delta<\delta_0.$ 

We also note that there is an obvious connection between $2$-LRC codes and
low-density parity-check codes whose graphs do not have cycles of length 4. While we employ some ideas from LDPC 
codes for the derivation of GV-type bounds, direct application of bounds on LDPC codes does not lead to good results
for the LRC problem.

{Existence of $t$-LRC codes with arbitrary $t$ and $r$ seems to be a difficult problem. We observe that there
is a connection between local recovery and expansion properties of some graph related to the code. The best
known expanders are constructed using the probabilistic method. Using them, we are able to show that there
exist asymptotically good $q$-ary $t$-LRC codes for any $t$ and $r$ over alphabets of large size $q$.

The version of LRC codes considered above assumes that every coordinate of the code can be recovered from
a few other coordinates. A less restricted version of this definition requires that this property applies only
to information symbols of the codeword. Accordingly, the two versions of LRC codes are called codes with
all-symbol locality and codes with information locality. In this paper we consider only the first of these
possibilities. Codes with multiple disjoint repair groups under the information locality assumption were
considered in \cite{wang2014a,raw14}.

Finally, we mention some other extensions and generalizations of the locality problem. In \cite{ankit-arya}, 
the notion of codes with locality was generalized to codes that enable
{\em cooperative} recovery from multiple erasures. In particular, this paper studied LRC codes that
support recovery of any $l$ failed codes symbols by reading at most $r$ other code symbols. A related paper
\cite{pra14} studied codes that enable {\em successive} local recovery of two erasures performed using
two recovering sets one after the other.

\section{An overview of bounds for LRC codes}
Below we give a brief overview of the known bounds on LRC codes with all-symbol locality.

\subsection{Known results, Single recovering set}
Clearly, any upper bound on the cardinality of a code with a given distance applies to LRC codes as well. We
are interested in bounds that in addition take account of the locality constraint.

Let $\cC$ be an $(n,k,r)$ LRC code.  
The rate of $\cC$ satisfies
				\begin{equation}
							\frac{k}{n}\leq  \frac{r}{r+1}.
							\label{prop546}
				\end{equation}
				The minimum distance of $\cC$ 
				satisfies 
   			\begin{equation}
				d\leq n-k-\ceilenv{ \frac{k}{r}}+2.				 
				\label{eq:sb}
     		\end{equation}
These upper bounds on the distance and rate of LRC codes were proved in \cite{Gop11,Pap12}.
The bound \eqref{eq:sb} forms a generalization of the classical Singleton bound in coding theory \cite{mac91}, 
and reduces to it for the maximum value of locality $r=k.$ Recently codes that generalize Reed-Solomon codes and achieve the bound 
\eqref{eq:sb} for small code alphabets and any $n$ a multiple of $r+1$ were constructed in \cite{Tamo13}.

The bound \eqref{eq:sb} does not account for the size of the code alphabet $q$. A {\em shortening bound} on the distance that 
depends on $q$ was derived in \cite{Cad13a}. To introduce it, denote by $M_q(n,d)$ the maximum cardinality of
a code in the $q$-ary Hamming space with distance $d$ and let $k_q(n,d):=\log_q M_q(n,d)$. For any 
$q$-ary LRC code with the parameters $(n,k,r)$ and distance $d,$
  \begin{equation}\label{eq:cm}
    k\le \min_{1\le s\le\min(\lceil\frac n{r+1}\rceil,\lceil\frac kr\rceil)} \,\{sr+k_q(n-s(r+1),d)\}.
  \end{equation}
Turning to asymptotic bounds, let us introduce the notation
   \begin{equation}\label{eq:rate1}
   R_q(r,\delta)=\limsup_{n\to\infty}\frac1n\log_q M_q(n,r,\delta n)
   \end{equation}
where $M_q(n,r,d)$ is the maximum cardinality of the code of length $n$, distance $d$, and locality $r$.
\begin{proposition}{\rm(\cite{Cad13a})} 
The following asymptotic bounds on the rate of $q$-ary codes with a single recovering set and locality $r$ hold true:
   \begin{align}
   R_q(r,\delta)&\le \frac r{r+1}(1-\delta), \quad 0\le\delta\le 1 \label{eq:Singleton}\\
   R_q(r,\delta)&\le \frac r{r+1}\Big(1-\delta\frac{q}{q-1}\Big), \quad 0\le\delta\le q/(q-1)\label{eq:Plotkin}\\
   R_q(r,\delta)&\le \min_{0\le \tau\le \frac 1{r+1}}\!\!\Big\{\tau r+(1-\tau(r+1))f_q\Big(\frac\delta{1-\tau(r+1)}\Big)\Big\}\label{eq:MRRW}    \end{align}
where    
  \begin{gather*}
  f_q(x):=h_q({\textstyle \frac 1q}(q-1-x(q-2)-2\sqrt{(q-1)x(1-x)})), \nonumber \\
         h_q(x):=-x\log_q(x/(q-1))-(1-x)\log_q(1-x). \nonumber
  \end{gather*}
\end{proposition}
Bounds \eqref{eq:Singleton}, \eqref{eq:Plotkin}, \eqref{eq:MRRW} follow on substituting into \eqref{eq:cm} classical upper bounds on $M(n,d)$. Namely \eqref{eq:Singleton} and \eqref{eq:Plotkin} are obtained using the Singleton
and Plotkin bounds, respectively, \cite{mac91}, while \eqref{eq:MRRW} follows on substituting the linear programming bound for $q$-ary codes \cite{aal77} (bound \eqref{eq:Singleton} can be also obtained by passing to the 
limit $n\to\infty$ in \eqref{eq:sb}).  For small values of $\delta$ a bound  slightly better than \eqref{eq:MRRW} can be obtained by using in \eqref{eq:cm} a better linear programming bound from \cite{aal90}. 

In the binary case, the bounds \eqref{eq:Singleton}-\eqref{eq:MRRW} are shown in Fig.~\ref{fig1}(a) below together
with a GV-type bound \eqref{eq0:gv} derived in this paper; see Theorem B and Theorem \ref{thm:gv1}.
It is evident from the plot that bound \eqref{eq:MRRW} changes its behavior for large values of $\delta.$ 
The reason for this is that when $\tau\to 0,$ the right-hand side of \eqref{eq:MRRW} approaches the value $f_q(\delta).$ At this point in the plot we switch to the classical (i.e., locality-unaware) linear programming bound
on the rate of codes $R_q(\delta)\le f_q(\delta).$

The following proposition follows by concatenating several copies of a single parity check code.
\begin{proposition}\label{prop:0} $R_q(r,0)\ge r/(r+1).$\end{proposition}

Other upper bounds on the distance of LRC codes appear in  \cite{pra14, wang2014b}. In particular,  \cite{wang2014b} 
gives an integer-programming based bound on the distance of LRC codes. The result of this 
paper is not expressed in a closed form, but is shown to improve the known bounds in many examples. 
In the case of linear cyclic LRC codes there is an obvious link between locality and the dual distance of the code, which enables one to use linear programming bounds on the code parameters. More details about this are given in \cite{tamo15b}.

As for lower bounds, the following results appear in the literature. By a straightforward adaptation of 
the GV argument \cite{Cad13a,barg15a} one obtains the following proposition.
\begin{proposition} A linear $(n,k,r)$ LRC code with distance $d$ exists if
    \begin{equation}\label{eq:dgv}
       \sum_{i=0}^{d-2}\binom{n-1}i(q-1)^i< q^{n-k-\lceil \frac n{r+1}\rceil}.
    \end{equation}
\end{proposition}
Below in \eqref{eq0:gv} we establish a more accurate version of the GV bound relying on the ideas from bipartite-graph and
LDPC codes.

Constructions of LRC codes on algebraic curves were recently proposed in \cite{barg15a}. 
Using asymptotically maximal curves, it is possible to construct sequences of LRC codes 
that improve the GV-type bound \eqref{eq0:gv}. The cardinality of the alphabet $q$ for which the
improvement takes place depends on the value of locality $r.$ For instance,
for $r=2$ it is possible to construct codes that asymptotically exceed the GV bound \eqref{eq0:gv} for alphabets of size $q\ge 289,$ and for $r=3$ for $q\ge 361.$

A more general version of LRC codes was introduced in \cite{kamath2013} which suggested considering codes whose coordinates can be partitioned into local codes which are $(r+\rho,r)$ MDS codes, for $\rho\ge 2$.  
In our terms this extended definition implies that for every
$\rho$-tuple of coordinates $i_1,\dots, i_\rho$ within the same local code there is a subset $\cR,|\cR|=r$ such that the symbols 
$x_{i_1},\dots,x_{i_\rho}$ 
of every codeword of the code can be reconstructed from the restriction of this codeword to the coordinates in $\cR$. A generalization of the bound \eqref{eq:sb} for this type of codes was obtained in \cite{kamath2013}.
Structural properties and existence of codes attaining this bound were considered in \cite{song14}, while an
algebraic construction of codes that attain this bound was proposed in \cite{Tamo13}.

\subsection{Known results, Multiple recovering sets}
The following bound on the distance of an $(n,k,r,t)$ LRC code was proved in \cite{raw14,wang2014a}
    \begin{equation}\label{eq:rpdv}
   d\le n-k+2-\Big\lceil \frac{t(k-1)+1}{t(r-1)+1}\Big\rceil.
   \end{equation}
This result is a direct generalization of the bound \eqref{eq:sb} and extends
the argument in \cite{Gop11} from one to many recovering sets.

Turning to lower bounds on the cardinality of codes with multiple recovering sets, let us first assume that $t=2.$
As observed in \cite{Tamo13,Tamo14,ankit-arya}, a natural way to construct codes with two recovering sets arises by 
using two-level code constructions such as product codes or codes on bipartite graphs. For instance, consider the case of graph 
codes, and take the example of a code on a bipartite regular graph where the edges incident to every vertex
form a codeword of the $[7,4,3]$ Hamming code $H_3$. Clearly, every coordinate can be recovered from a parity check of
4 symbols in two independent ways since the dual code $H_3^\bot$ has distance 4. In other words, we obtain an
$(n,k,3,2)$ LRC code. It is possible to estimate the dimension and distance of the resulting code \cite{bar06c}, 
and we obtain a family of asymptotically good $2$-LRC codes. Many more examples can be constructed using this general
approach. For greater $t$, similar constructions can be obtained using codes on regular hypergraphs \cite{bar08b}. 

To give a simple example of multilevel constructions, consider a product code formed of two single parity check codes
with $r$ message symbols each. The rate of the resulting code equals $r^2/(r+1)^2,$ and each symbol has locality $r$.
Generalizing, we can construct a $t$-th power of the binary $(r+1,r)$ single parity check code and obtain a code with $t$ disjoint 
recovering sets that has the rate $(r/(r+1))^t.$ 

  Define
  \begin{align}\label{eq:rt}
   R^{(t)}_q(r,\delta)&=\limsup_{n\to\infty}\frac1n\log_2 M^{(t)}_q(n,r,\delta n)
  \end{align}
where $M^{(t)}_q(n,r,d)$ is the maximum cardinality of the $q$-ary code of length $n$, distance $d$, and
$t$ recovering sets of size at most $r$ for every symbol.
In the particular case of $t=1$ the quantity
$R^{(t)}_q(r,\delta)$ is the same as the function $R_q(r,\delta)$ defined in \eqref{eq:rate1} above. Clearly, $R^{(t)}_q(r,\delta)\le R_q(r,\delta),$ so all the upper bounds of the previous section apply to
the current case. 
\remove{, and where $\delta^{(t)}_q(r,R)$ is the maximum distance of a code
of size $q^{Rn}$ with $t$ recovering sets\footnote{The upper bounds are easier expressed using the quantity
\eqref{eq:dt}, while for the lower bounds the more convenient of the two is \eqref{eq:rt}.}}

From \eqref{eq:rpdv} we obtain
  \begin{equation}\label{eq:at1}
  R_q^{(t)}(r,\delta)\le \frac{t(r-1)+1}{tr+1}(1-\delta), \quad0\le\delta\le 1.   
   \end{equation}
Below in \eqref{eq:sa} we will obtain a somewhat tighter asymptotic bound on $R^{(t)}_q.$

Algebraic constructions of LRC codes with $t\ge 2$ recovering sets were considered in \cite{Tamo13,barg15a}.
Block designs were used in \cite{AnyuWang} to construct binary $t$-LRC codes for any $r$ and $t$,
 resulting in codes of rate $R=\frac{r}{r+t}$ and minimum distance $d=t+1$. 
 
\subsection{New bounds on LRC codes}
In this section we summarize the main contributions of this paper.

\vspace{5mm}\noindent{\bf Theorem A}.
{\em Let $\cC$ be an $(n,k,r,t)$ LRC code with $t$ disjoint recovering sets of size $r$. Then the rate of $\cC$ satisfies
\begin{equation}
\frac{k}{n}\leq \frac{1}{\prod_{j=1}^t(1+\frac{1}{jr})}.
\label{eq:rate}
\end{equation}
The minimum distance of $\cC$ is bounded above as follows:
   \begin{align} 
  d&\leq  n-\sum_{i=0}^t\Big\lfloor\frac{k-1}{r^i}\Big\rfloor.       \label{eq:d2}
         \end{align}
   }
 {\sc Remarks:} \\ 
 \remove{1. For $t=1$ the bound on the rate \eqref{cvcv} reduces to \eqref{prop546}. 
For general $t$ the expression 
on the right-hand side of \eqref{cvcv} approximately equals $t^{-1/r}$ (more precise results
 are established below in the paper; see Lemma \ref{lemma:rroot}). }
 \nd 1. For codes with a single recovering set for every symbol, the bound on the rate \eqref{eq:rate} reduces to \eqref{prop546}, which is a tight bound \cite{Tamo13}. Currently $t=1$ is the only case for which the known
 bounds on the rate have been shown to be tight. For two recovering sets the bound \eqref{eq:rate} takes the form 
   \begin{equation}\label{eq:r2r}
   \frac{k}{n}\leq \frac{2r^2}{(r+1)(2r+1)}.
   \end{equation}
At the same time, the product construction mentioned above gives $\frac kn=r^2/(r+1)^2$ which is only slightly less than \eqref{eq:r2r}.

\vspace*{.05in}\nd 2.  For codes with a single recovering set for every symbol, the bound on the distance \eqref{eq:d2} reduces to \eqref{eq:sb}, and there exist large families of codes that meet this bound with equality \cite{Tamo13,sil13,matroid-paper}.
The next interesting case, in particular for applications, is $t=2.$  From \eqref{eq:d2} we obtain the bound
  \begin{equation}\label{2}
  d\leq n-\Big( k-1+\Big \lfloor \frac{k-1}{r}\Big\rfloor+\Big\lfloor \frac{k-1}{r^2}\Big\rfloor\Big).
  \end{equation}
For some parameters this bound is also tight. For instance, consider the shortened binary Hamming code of length $6$ with
the parity-check matrix

 {\small $$
  \begin{pmatrix}
  0&0&0&1&1&1\\
  0&1&1&0&0&1\\
  1&0&1&0&1&0
  \end{pmatrix}.
  $$}
\hspace*{-.1in} It is easily seen that this is a $(6,3,2,2)$ LRC code, and its distance $d=3$ meets the bound \eqref{2} with equality.

\vspace{5mm}\noindent{\bf Corollary}.
{\em The rate of an $(n,k,r,t)$ LRC code satisfies
     \begin{equation}
       \frac kn \le \frac 1{\sqrt[r]{t+1}}.
       \label{eq:er}
       \end{equation}
For any alphabet size $q$ the following asymptotic bound holds true:
   \begin{align}
     R_q^{(t)}(r,\delta)&\le \frac{r^t(r-1)}{r^{t+1}-1}(1-\delta), \quad 0\le\delta\le 1.\label{eq:sa}     
    \end{align}
The bound \eqref{eq:sa}
is tighter than the asymptotic version of the bound \eqref{eq:rpdv}  given in \eqref{eq:at1} for all $R$.

We also have 
  \begin{gather*}
  R_q^{(t)}(r,\delta)=0, \quad \frac{q-1}q\le \delta\le 1\\
  \frac{r}{r+t}\le R_q^{(t)}(r,0)\le \frac{r^t(r-1)}{r^{t+1}-1}.
  \end{gather*}
} 
\begin{IEEEproof} The estimate \eqref{eq:er} follows from Lemma \ref{lemma:rroot} proved below in Sect.~\ref{sect:rb}.
 Estimate \eqref{eq:sa} follows immediately from \eqref{eq:d2}. 

Let us show that \eqref{eq:sa} is a tighter bound than \eqref{eq:at1}. We have
  \begin{align*}
   \frac{r^{t+1}-r^t}{r^{t+1}-1}-\frac{t(r-1)+1}{tr+1}=\frac{t(r-1)-r^t+1}{(r^{t+1}-1)(tr+1)}.
    \end{align*}
The numerator of the last fraction is negative for all $r,t\ge 2,$ which is easy to see, for instance, by induction
on $r$. This proves our claim. 

The result for large $\delta$ is implied by the Plotkin bound. The lower bound on $R_q^{(t)}(r,0)$
follows from the result of \cite{AnyuWang} mentioned above.

  \end{IEEEproof} 

\vspace{.1in}\noindent{\bf Theorem B}. {\em    The following asymptotic GV-type bounds for LRC codes hold true:
\begin{multline}\label{eq0:gv}
    R_q(r,\delta)\ge 1 - \min\limits_{0<s\le 1} 
    \Big\{ \frac{1}{r+1}\log_q ( ( 1 + (q-1)s)^{r+1} \\+ (q-1) (1-s)^{r+1}) - \delta \log_q s \Big\}.
    \end{multline}
       \begin{equation}\label{eq0:gv2}
R^{(2)}_q(r,\delta) \ge \frac{r}{r+2} - \min_{0<s\le 1} 
\Big\{ \frac{1}{\binom{r+2}{2}}\log_q g_q^{(2)}(s) - \delta \log_q s \Big\},
   \end{equation}
where $g_q^{(2)}(s)$ is given in \eqref{eq:local}-\eqref{eq:E} below. In particular, for $q=2$ 
       \begin{multline}\label{eq0:local}
   g_2^{(2)}(s)=\frac{1}{2^{r+2}} \sum\limits_{i = 0}^{r+2} \binom{r+2}{i} (1+s)^{\binom{r+2}{2} - i(r+2-i)}\\
   \times (1-s)^{i(r+2-i)}. 
   \end{multline}
We also have $R_q(r,\delta)>0,0\le\delta<(q-1)/q,$
     \begin{gather}
   R_q(r,0)=\frac{r}{r+1}, \quad R_q(r,\delta)=0, \;{\textstyle \frac{q-1}q}\le\delta\le 1 \label{eq:pb1}\\
  \frac r{r+2}\le R_q^{(2)}(r,0)\le \frac{r^2(r-1)}{r^{3}-1}\\ R_q^{(2)}(r,\delta)=0, \;{\textstyle \frac{q-1}q}\le\delta\le 1.
  \label{eq:pbm}
   \end{gather}
   } 
 Independently, the bound \eqref{eq0:gv} was obtained in \cite{Cad13a}. 
   
In Fig.~\ref{fig1} below we show the GV-type bounds for LRC codes. In Fig.~\ref{fig1}(a) 
the bound \eqref{eq0:gv} is plotted together with the
upper bounds \eqref{eq:Singleton}-\eqref{eq:MRRW}. In Fig.~\ref{fig1}(b) the bound \eqref{eq0:gv2} is
plotted together with other lower and upper bounds.
 
The question of lower bounds for $t\ge 3$ recovering sets becomes more difficult because of the complicated nature of 
interaction between the sets. Using graph-theoretic arguments, we establish the following asymptotic result.
 
\vspace{.1in}\noindent{\bf Theorem C}. 
{\em  
For sufficiently large $q,$ there exists a sequence of $t$-LRC codes with locality $r\ge t$ and rate
 $R, 0\le R\le 1-t/(r+1)$ such that the relative distance $\delta$ is determined from the following
 two equations with respect to the unknowns $\delta,\gamma:$ 
\begin{align}
\delta(1-t\gamma)=1- &\frac{t}{r+1}-R
\label{eq:0}\\
\frac{t-1}{t} h(\delta)-\frac{1}{r+1} & h (\delta\gamma(r+1))\nonumber\\
-&\delta\gamma(r+1) h\Big(\frac{1}{\gamma(r+1)}\Big)=0,
\label{eq:1}
\end{align}
where $1/(r+1)\le \gamma\le1/t,$ and $h=h_2$ is the binary entropy function (viz.~\eqref{eq:MRRW}).}
 
\vspace*{.1in}
In Fig.~\ref{fig1}(c) below the bound of this theorem is plotted together with the Singleton-type bound of \eqref{eq:sa}.

\vspace*{.1in}
{\em Remarks.} 1. {\em Corner points:} The pair $(R,\delta)=(1-t/(r+1),0)$ provides a trivial
solution to \eqref{eq:0}-\eqref{eq:1}, accounting for one of the two endpoints of the lower bound. At
the same time, the pair $(\gamma,\delta)=(1/(r+1),1)$ satisfies \eqref{eq:0}-\eqref{eq:1}, resulting in
$(R,\delta)=(0,1).$

2. For small values of $\delta,$ Equations \eqref{eq:0},\,\eqref{eq:1} do not have a solution for $\gamma$ in the segment $[1/(r+1),1/t].$
This corresponds to $\delta$ smaller than the value $\delta^\ast$ that gives the maximum possible rate of $R=1-t/(r+1)$ according to
\eqref{eq:0}.
For $\delta<\delta^\ast$ the best we can claim is the existence of codes of the rate $R=1-t/(r+1),$ extending the
bound by a horizontal tangent line. This setting is close to the problem of locally decodable codes \cite{yek12}, where we do not
attempt to construct codes with some particular distance, focusing instead on the local decoding property and high rate.
Note that finding maximum rate of 
locally decodable  codes currently is an open problem.

3. \emph{The Singleton bound}. The Singleton bound on the distance of codes without the locality constraint is
attained by Reed-Solomon codes for all values of the distance. Allowing the alphabet size to increase with $n$, we 
see that this bound is also asymptotically tight for $n\to \infty.$ The same conclusion is true for LRC codes
with a single recovering set because of the construction of \cite{Tamo13}. At the same time, for
codes with $t\ge 2$ recovering sets there is a gap between the best known lower and upper bounds on 
$R^{(t)}_q(r,\delta)$ for all values of the alphabet size. It is not clear at this point, which of the two bounds in
\eqref{eq:0}-\eqref{eq:1} and \eqref{eq:sa} is loose, and it is possible that both can be improved. 
 
4. \emph{ The case of $t=2$ recovering sets:} The simplest unresolved case and in some sense the most interesting one for applications that require high availability is the case of two recovering sets. Paper \cite{Tamo13} 
gave an explicit construction of $2$-LRC codes with relative distance 
$$\delta\geq 1-R\frac{r+1}{r-1}$$
for any given value of $q$.  
We claim that for large alphabets the result of Theorem C improves upon this bound.
Indeed, for a given $R$ the curves in the $(\gamma,\delta)$ plane defined in  \eqref{eq:0} and \eqref{eq:1} intersect for $\gamma>\frac{1}{r+1}$ (see the proof of Lemma \ref{existence}). Moreover, by \eqref{eq:0}  $\delta=\delta(\gamma)$ is a strictly increasing function 
that takes the value $1-R\frac{r+1}{r-1}$ for $\gamma=1/(r+1).$ Hence we conclude that
Theorem C establishes existence of a sequence of $2$-LRC codes with higher minimum distance than \cite{Tamo13}.


\begin{figure*}[ht] \vspace*{.1in}\centering
\subfigure[Bounds for codes with single recovering set.
]{\includegraphics[scale=0.56]{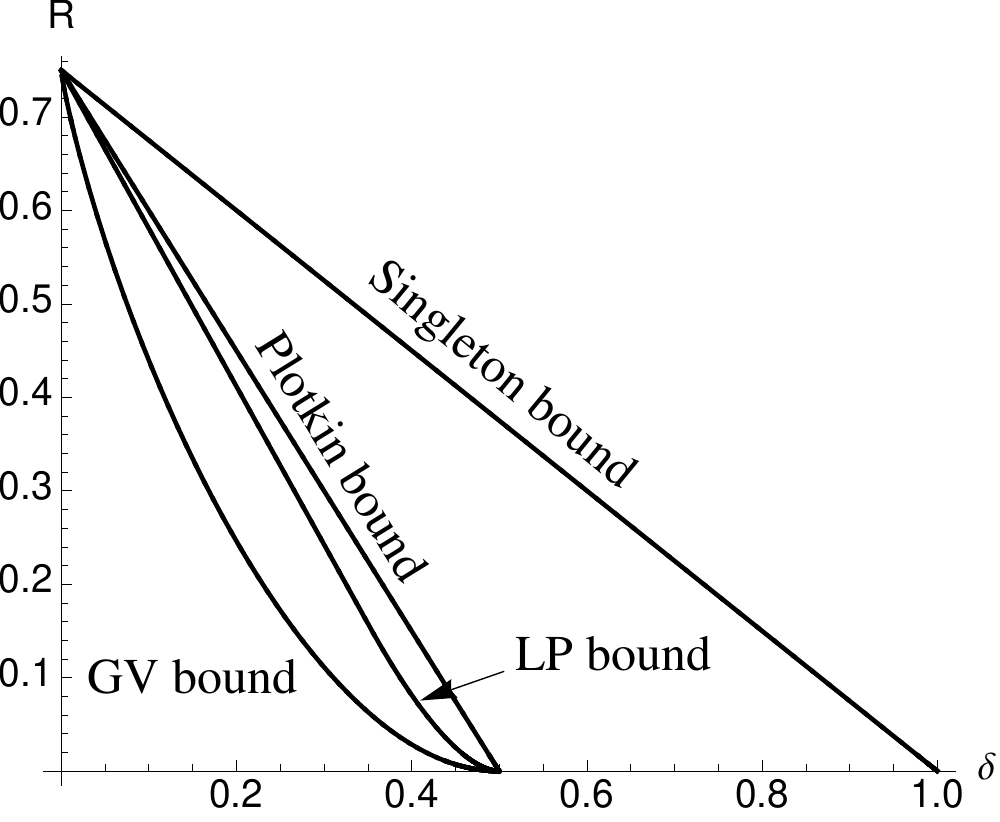}} \label{fig1:a}
\hspace*{.2in}\subfigure[Asymptotic bounds on codes with one and two recovering sets.]{\includegraphics[scale=0.52]{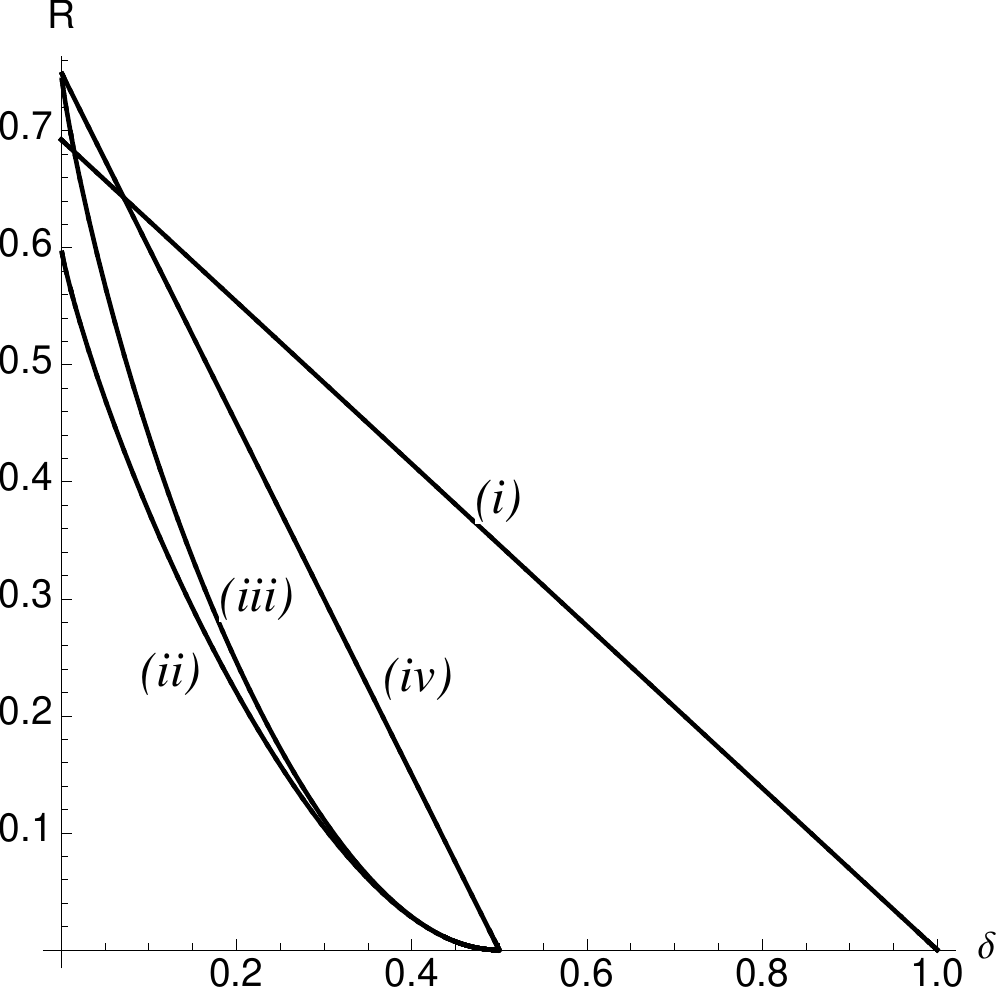}}\label{fig1:b}
\hspace*{.2in}\subfigure[Bounds for $t=3$ recovering sets (large $q$)]{\includegraphics[scale=0.55]{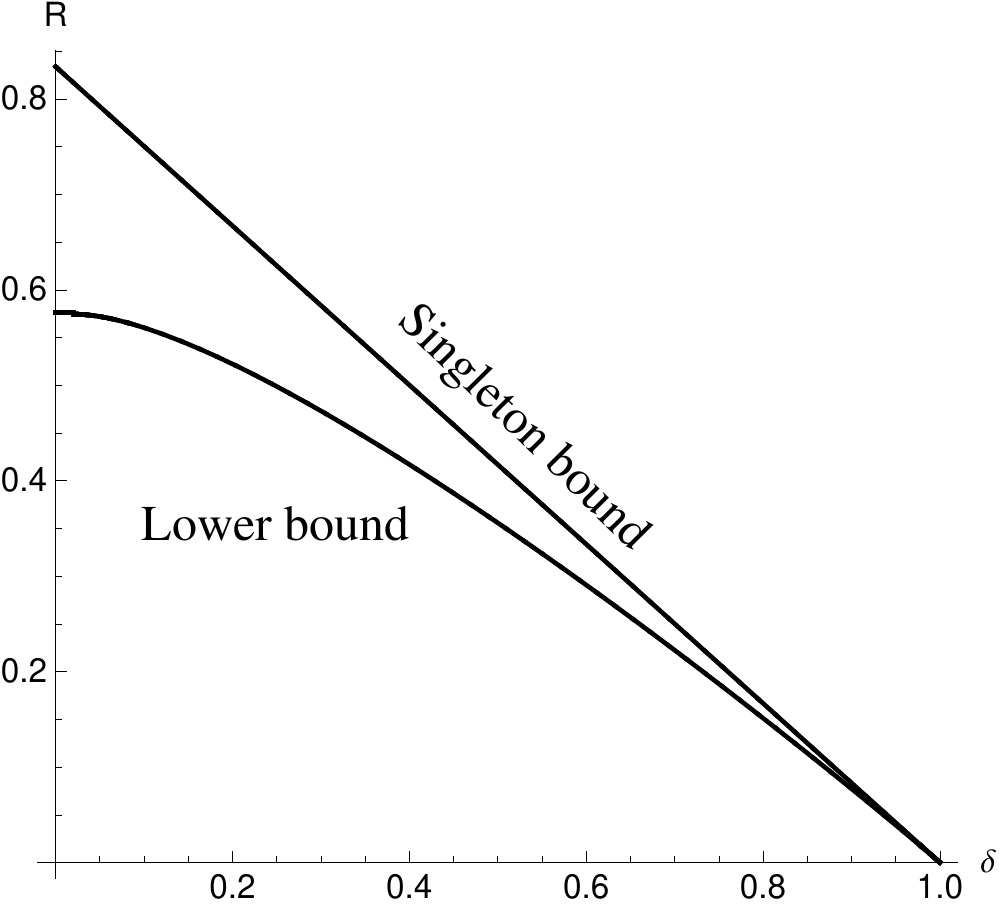}}
\caption[caption]{Asymptotic bounds for LRC codes. 
\\
(a) Binary codes, $r=3$. The plot shows the GV-type bound \eqref{eq0:gv} and upper bounds \eqref{eq:Singleton} (the Singleton bound), \eqref{eq:Plotkin} (the Plotkin bound), \eqref{eq:MRRW}
(the Linear Programming bound).

(b) Asymptotic upper and lower bounds for binary codes with one and two recovering sets ($t=1,2\;\,r=3$). Plot $(i)$ shows the Singleton-type bound \eqref{eq:sa}. The curve marked $(ii)$ is the asymptotic GV-type bound on $R^{(2)}_2(r,\delta).$ For reference we also show $(iii)$ the GV and $(iv)$ Plotkin bounds for codes with one recovering set (copied from part (a)). Note that $R_2(r,\half)=R^{(2)}_2(r,\half)=0$. 

(c) Bounds for LRC codes with $r=6$ and $t=3$ recovering sets (large $q$). The plot shows the lower bound \eqref{eq:0}-\eqref{eq:1} together with the Singleton-type bound \eqref{eq:sa}.
}\label{fig1}
  \end{figure*}

 \remove{
\begin{figure}[ht] \vspace*{.2in}\centering{\includegraphics[scale=0.65]{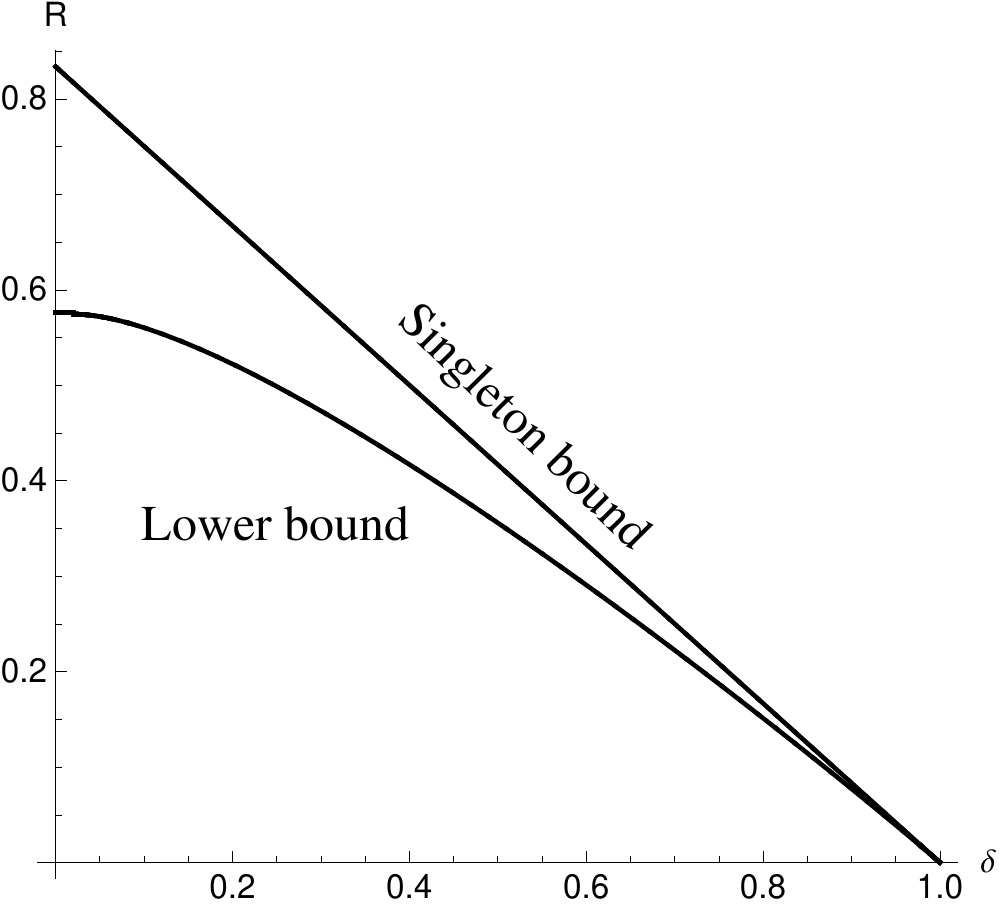}} 
\centering\caption{Lower bound for LRC codes with $r=4$ and $t=2$ recovering sets (large $q$), shown together with the Singleton-type bound \eqref{eq:sa}.} \label{fig2}
  \end{figure}}

\section{An Upper Bound on the Rate of LRC Codes}
\label{sect:rb}
In this section we prove estimate \eqref{eq:rate} in Theorem A.
\subsection{The recovery graph}
\label{the recovering graph}
Assume that coordinate $i$ has $t$ disjoint recovering sets $\cR_i^1,...\cR_i^t$,  each of size $r$, 
where $\cR_i^j\subset [n]\backslash i$.
Define a directed graph $G$ as follows. The set of vertices $V=[n]$ corresponds to the set of $n$ coordinates of the LRC code. The ordered pair of vertices $(i,j)$ forms a directed edge $i\to j$ if $j\in {\cR}_i^l$ for some $l\in[t].$ 
We color the edges of the graph with $t$ distinct colors in order to differentiate between the recovering sets of each coordinate. 
More precisely, let $F_e:E(G)\to[t]$ be a coloring function of the edges,
given by $F((i,j))=l$ if $j\in {\cR}_i^l.$ 
Thus, the out-degree of each vertex $i\in V=V(G)$ is $\sum_{l}|{\cR}_i^l|=tr,$ and the edges leaving $i$ are colored in 
$t$ colors. We call $G$ the {\em recovery graph} of the code $\cC.$

The following lemma will be used in the proof.
\begin{lemma} \label{lemma:rfrf} 
There exists a subset of vertices $U\subseteq V$ of size at least 
  \begin{equation}\label{eq:size}
  |U|\geq n\Big(1-\frac{1}{\prod_{j=1}^t(1+\frac{1}{jr})}\Big)
  \end{equation}
such that for any $U'\subseteq U$, the induced subgraph $G_{U'}$ on the vertices $U'$ 
has at least one vertex $v\in U'$ such that its set of outgoing edges $\{(v,j),j\in U')\}$ is missing at least one color. 
\end{lemma}

\begin{IEEEproof}
For a given permutation $\tau$ of 
the set of vertices $V=[n]$, we define the coloring of some of the vertices as follows:  
The color $j\in [t]$ is assigned to the vertex $v$ if 
   \begin{equation}
     \tau(v)>\tau(m) \quad\text{for all }m\in {\cR}_v^j.
   \label{eq:dfdf}
   \end{equation}
If this condition is satisfied for several recovering sets ${\cR}_v^j$, the vertex $v$ is assigned any of the colors $j$ corresponding to these sets. 
Finally, if this condition is not satisfied at all, then the vertex $v$ is not colored.

Let $U$ be the set of colored vertices, and consider one of its subsets $U'\subseteq U$.
\remove{For a color $i,1\le i\le t$ and a vertex $v$ denote by 
$N_i(v)$ the set of vertices $u$ such that $F_e((v,u))=i,$ i.e., $u\in {\cR}^i_v.$

Let $\tau$ be some permutation on $n$ elements and let $U$ be the set of all the vertices $v\in V$ such that there exists $i\in[t]$ with
the property that $\tau(v)$ is greater than
all the values $\tau(u), u\in N_i(v),$ i.e., 
  \begin{equation}\label{eq:ti}
   \tau(v)> \max_{u\in N_i(v)}\tau(u).
  \end{equation}
Pick a random permutation $\tau$ for the set of vertices $V$, and define the set $U$ to contain all the vertices $v$ such that $\tau(v)$ is greater than the values given by $\tau$ to at least one color set $N_i(v),$ namely there exists 
$i$ such that 
Next, color the vertices of $U$ using the $m$ colors in the following manner: A vertex $v$ in $U$ will be colored in color $i$ if 
\begin{equation}
\tau(v)>\max_{u\in N_i(v)}\tau(u).
\label{eq:rdrd}
\end{equation}
Note that if (\ref{eq:rdrd}) is satisfied for several colors, then color the vertex $v$ arbitrarily with one of these colors. By the definition of the set $U$, it is clear that this defines a coloring for all the vertices of $U$. We claim that the set $U$ is the desired set.} 
Let $G_{U'}$ be the induced subgraph on $U'$. We claim that there exists $v\in U'$ such that its set of outgoing edges is missing at least one color in $G_{U'}$. Assume toward a contradiction that every vertex 
of $G_{U'}$ has outgoing edges of all $t$ colors. 
Choose a vertex $v\in U'$ and construct a walk through 
the vertices of $G_{U'}$ according to the following rule. 
If the path constructed so far ends at some vertex with color $j,$ choose one of its outgoing edges also
 colored in $j$ and leave the vertex moving along this edge. 
 By assumption, every vertex has outgoing edges of all $t$ colors, so this process, and hence this path can be extended indefinitely. 
Since the graph $G_{U'}$ is finite, there will be a vertex, call it $v_1,$ that is encountered twice. 
The segment of the path that begins at $v_1$ and returns to it has the form
   $$
   v_1\rightarrow v_2\rightarrow... \rightarrow v_l,
   $$
where $v_1=v_l$. For any $i=1,...,l-1$ the vertex $v_i$ and the edge $(v_i,v_{i+1})$ are colored with 
the same color. Hence by the definition of the set $U$ we conclude that 
$\tau(v_i)>\tau(v_{i+1})$ for all $i=1,\dots, l-1,$ a contradiction. 

In order to show that there exists such a set $U$ of large cardinality, we choose the permutation
$\tau$ randomly and uniformly among all the $n!$ possibilities and compute the expected cardinality 
of the set $U.$

Let $A_{v,j}$ be the event that \eqref{eq:dfdf} holds for the vertex $v$ and the color $j.$ Since $\Pr(A_{v,j})$ does not depend
on $v$, we suppress the subscript $v$, and write
  $$
  \Pr(v\in U)=\Pr( \cup_{j=1}^t A_{j}).
  $$
Let us compute the probability of the event $\cup_{j=1}^tA_j.$
Note that for any set $S\subseteq [t]$ the probability of the event that all the $A_j,j\in S$ occur simultaneously, equals 
     $$
     P(\cap_{j\in S}A_j)=\frac{1}{|S|r+1}.
     $$
Hence by the inclusion exclusion formula we get 
\begin{align}
\Pr(\cup_{j=1}^t A_j)&=\sum_{j=1}^t(-1)^{j-1}\binom{t}{j}P(A_1\cap...\cap A_j)\nonumber\\
&=\sum_{j=1}^t(-1)^{j-1}\binom{t}{j}\frac{1}{jr+1}\nonumber\\
&=-\frac{1}{r}\Big(\sum_{j=0}^t(-1)^j \binom{t}{j}\frac{1}{j+\frac{1}{r}}-r\Big)\nonumber\\
&=1-\frac{1}{r}\sum_{j=0}^t(-1)^j \binom{t}{j}\frac{1}{j+\frac{1}{r}}\nonumber\\
&=1-\frac{1}{r}\frac{t!}{\frac{1}{r}(1+\frac{1}{r})...(t+\frac{1}{r})}\label{titi}\\
&=1-\frac{1}{\prod_{j=1}^t(1+\frac{1}{jr})}\nonumber,
\end{align}
where \eqref{titi} follows from \cite[p.~188]{graham88}.
Now let $X_v$ be the indicator random variable for the event that $v\in U$, then
\begin{align*}
\E(|U|)&=\sum_{v\in V}\E(X_v)\\
&=\sum_{v\in V}\Pr(v\in U)\\
&=n\Pr( \cup_{j=1}^t A_{j})\\
&=n(1-\frac{1}{\prod_{j=1}^t(1+\frac{1}{jr})}).
\end{align*}
The proof is completed by observing that there exists at least one choice of $\tau$ for which $|U|\geq \E(|U|).$
\end{IEEEproof}
\remove{\begin{theorem}
\label{thm1}
The rate of an $(n,k)$ LRC code with $t$ disjoint recovering sets, each of size at most $r$, satisfies
\begin{equation}
\frac{k}{n}\leq \frac{1}{\prod_{j=1}^t(1+\frac{1}{jr})}.
\label{cvcv}
\end{equation}}

\subsection{ Proof of the bound on the rate \eqref{eq:rate}}
Let $U\subseteq [n]$ be the set of vertices of cardinality as in \eqref{eq:size} constructed in Lemma \ref{lemma:rfrf} and let $\overline{U}=[n]\backslash U$ be its complement in $[n].$ 
We claim that the value of every coordinate $i\in U$ can be recovered by accessing the coordinates 
in $\overline{U}.$ To show this, we construct the following iterative procedure, which in each step is applied to the subset $U'\subseteq U$
formed of the coordinates whose values are still unknown.
In the first step $U'=U.$ By Lemma \ref{lemma:rfrf} the induced subgraph $G_{U'}$ contains a vertex $v\in U'$ that is missing one color, call it $i$. This means that the $i$-th recovering set of $v$
is entirely contained in $\overline{U'}$. Hence one can recover the value of the coordinate $v$ of the codeword
by 
knowing the values of the coordinates in $\overline{U'}$. In the next step use the same argument for the set 
of coordinates $U'\backslash\{v\}.$ In this way all the coordinates in $U$ are recovered step by step relying only on the values
of the coordinates in $\overline U.$ Therefore, 
    $$
    k\leq |\overline{U}|\leq \frac{n}{\prod_{j=1}^t(1+\frac{1}{jr})}
    $$
proving inequality \eqref{eq:rate}.

To get a clearer impression of \eqref{eq:rate}, observe that 
$$
\log \prod_{j=1}^t\Big(1+\frac1{jr}\Big)=\sum_{j=1}^t\log\Big(1+\frac1{jr}\Big)\approx \sum_{j=1}^t\frac1{jr}\approx\frac1r \log t.
$$
Therefore, the value of the product in \eqref{eq:rate} is about $\sqrt[r]t.$ More precisely, let us show
that \eqref{eq:rate} implies the bound \eqref{eq:er}.
\begin{lemma}\label{lemma:rroot}
$$\sqrt[r]{t+1}\leq \prod_{j=1}^t\Big(1+\frac{1}{jr}\Big)\leq\sqrt[r]{t+1}\Big(1+\frac{1}{r}\Big).$$
\end{lemma}

\begin{IEEEproof}
For $i=0,...,r-1$ define the quantity 
  $$
  f_i=\prod_{j=1}^t\Big(1+\frac{1}{i+jr}\Big).
  $$ 
It can be easily seen that for any $i$, 
   \begin{align}
     f_i\leq f_0 &\leq f_i\Big(1+\frac{1}{r}\Big)\Big(1+\frac{1}{(t+1)r}\Big)^{-1} \label{eq:1nnh} \\ \nonumber
     &=f_i\Big(1+\frac{t}{(t+1)r+1}\Big).\nonumber
\end{align} 
Furthermore
\begin{align}
\prod_{i=0}^{r-1}f_i&=\prod_{i=0}^{r-1}\prod_{j=1}^t(1+\frac{1}{i+jr})\nonumber\\
&=\prod_{j=r}^{(t+1)r-1}(1+\frac{1}{j})\nonumber\\
&=t+1\label{eq:2nnh}.
\end{align}

Using the inequalities \eqref{eq:1nnh} in \eqref{eq:2nnh}, we obtain 
\begin{align*}
\sqrt[r]{t+1}&=\sqrt[r]{\prod_{i=0}^{r-1}f_i}
\leq \sqrt[r]{\prod_{i=0}^{r-1}f_0}\\
&=\prod_{j=1}^t\Big(1+\frac{1}{jr}\Big)\\
&\leq \sqrt[r]{\prod_{i=0}^{r-1}f_i(1+\frac{t}{(t+1)r+1})}\\
&=\sqrt[r]{t+1}(1+\frac{t}{(t+1)r+1})\\
&\leq \sqrt[r]{t+1}(1+\frac{1}{r}).
\end{align*}
\end{IEEEproof}

\section{Upper Bounds on the Minimum Distance of LRC Codes}
In this Section we prove the bound \eqref{eq:d2} of Theorem A. Our approach extends the idea of \cite{Gop11} used to 
prove the bound \eqref{eq:sb}. We begin with a simple observation. Suppose we are given an LRC code $\cC$ 
with the parameters $(n,k,r,t)$ over a $q$-ary alphabet (field or not).
Let $I\subseteq [n]$ be a subset of coordinates, and let $x_I$ be a restriction to $I$ of a codeword $x\in \cC.$ Recall our notation
$C_I:=\{x_I: x\in \cC\}.$
%
Observe that if $I$ is such that $|\cC_I|<q^k$, then the distance of the code $\cC$ satisfies the inequality
 \begin{equation}\label{eq:dddd}
    d(\cC)\le n-|I|.
    \end{equation}
The main idea behind the proof of \eqref{eq:d2} 
is to construct a set $S$ of size $k-1$ such that the values of codeword coordinates in $S$ determine the values
of the coordinates in some large subset $S'$. Since $|\cC_{S\cup S'}|\le q^{k-1}$, we then can apply \eqref{eq:dddd} to derive
a bound on $d(\cC).$


\subsection{Proof of the bound \eqref{eq:d2} }
 Consider the recovery graph $G$ of an $(n,k,r,t)$ LRC code $\cC$ with $t$ recovering sets,
 defined in Sect.~\ref{the recovering graph}.
Consider the following coloring procedure of the vertices\footnote{
This coloring uses just one color (a vertex is colored or not) and is different from the $t$-coloring of the edges introduced in the beginning of Sect.~\ref{sect:rb}. Both colorings will be used in the proof}.
Start with an arbitrary subset of vertices
 $S\subseteq V$ and color it in some fixed color. Now let us color some of the remaining uncolored
 vertices according to the following rule. A vertex is colored if at least one of its recovering sets is completely colored.
This process continues until no more vertices can be colored (recall that $G$ is finite).
We denote the set of colored vertices obtained at this point by $\Cl(S)$ and call 
it the {\em closure} of $S$ in $G$. Call 
the quantity
$|\Cl(S)|/|S|$ the {\em expansion ratio} of the set $S.$
It is clear that a large expansion ratio means that 
the values of a large number of coordinates outside $S$ 
are determined by the values of the coordinates in $S.$
We shall show that there is a subset
with a large closure and use \eqref{eq:dddd} to bound the code's distance.

We begin with two lemmas.
\begin{lemma}
\label{best lemma}
Let $G$ be the recovery graph of an $(n,k,r,t)$ LRC code $\cC$. For any vertex $v\in G$ there exists a set $S$ of size at most $r^t$ such that $v\in \Cl(S),$ 
and the expansion ratio of $S$ is at least 
  \begin{equation}
e_t=\frac{r^{t+1}-1}{r^{t+1}-r^t}.
    \label{dikla6}
    \end{equation}
\end{lemma}
\begin{IEEEproof}
We use induction on $t$. For $t=0$ there are no edges in the graph. Define $S=\{v\}$ and note that $\Cl(S)=S=\{v\},$ 
and the expansion ratio is $1$ as needed. Now assume that the claim is correct for $t$ recovering sets. Let us prove it for 
$t+1$ recovering sets. Remove from $G$ a vertex $v$. For each other vertex $u\neq v$ we remove the edges that correspond 
to one of its recovering sets. Specifically, if $u$ has a recovering set that contains $v$, we
remove all of its edges that correspond to this recovering set; otherwise, remove the edges that correspond to any one of its recovering sets. 
Denote the resulting graph by $G_1$, and observe that each vertex of $G_1$ has exactly $t$ recovering sets. We will denote
by $\Cl_1(\cdot)$ the result of the closure operation in $G_1$ and use a similar notation for other graphs in the proof.

Let $v_1,...,v_l$ be the vertices of one of the recovering sets of $v$, where $l\leq r$. Our plan is to apply the induction hypothesis successively for each of the $l$ vertices, where in the $i$-th step we construct  a subset of vertices  $S_i\subseteq V(G_1)$ such that $v_i\in \Cl_1(S_1\cup...\cup S_i)$. Suppose that the subsets $S_1,...,S_{i-1}$ are already constructed. Color the vertices 
in $\Cl_1(S_1\cup ...\cup S_{i-1})$ and let $G_i$ be the induced subgraph of $G_1$ on the non-colored vertices of $G_1$, i.e., the set of vertices $V(G_i)=V(G_1)\backslash \Cl_1(S_1\cup ...\cup S_{i-1})$. 

Let us describe the construction of the set $S_i$.
If $v_i \in \Cl_1(S_1\cup ...\cup S_{i-1}),$ put $S_i=\emptyset.$ Otherwise $v_i\in V(G_i)$. 
Note that each vertex $u$ in $G_i$ has outgoing edges of all $t$ colors because otherwise, if $u$ is missing one color, 
it has a recovering set that is contained in $\Cl_1(S_1\cup ...\cup S_{i-1})$, and then also $u\in \Cl_1(S_1\cup ...\cup S_{i-1}).$ Hence $G_i$ can be viewed as a recovery graph of a code with $t$ recovering sets for each coordinate.
Apply the induction hypothesis to $G_i$ to find a set $S_i$ of size at most $r^t$ and expansion ratio at least $e_t$ 
such that $v_i\in \Cl_i(S_i),$ where $\Cl_i(\cdot)$ is the closure in $G_i.$ Notice that since $\Cl_i(S_i)$ is a subset of the vertices of the graph $G_i$, it is disjoint from the set 
$\Cl_1(S_1\cup...\cup S_{i-1})$. Furthermore, it is easy to see that
  \begin{align}
  \Cl_1(S_1\cup...\cup S_{i-1}\cup S_i)&=\Cl_1(\Cl_1 (S_1\cup...\cup S_{i-1}) \cup S_i) \nonumber\\
  &=\Cl_1(S_1\cup...\cup S_{i-1})\cup \Cl_i(S_i)\nonumber\\
  &=\bigcup_{j=1}^i\Cl_j(S_j), \label{eq:Cl}
  \end{align}
where the union is in fact a disjoint union. 
We claim that the set
   $
   S=\cup_{i=1}^lS_i
   $ 
satisfies the properties in the statement of the lemma.
%
%

We need to show that $v\in \Cl(S).$
First let us show that for any $i=1,...,l$ the vertex $v_i\in \Cl_1(S)$. 
Indeed, by construction, if $S_i$ is the empty set, then $v_i\in \Cl_1(S_1\cup...\cup S_{i-1})$, otherwise $v_i\in \Cl_i(S_i)$. 
We conclude that $\Cl_1(S)$ contains a complete recovering set $v_1,...,v_l$ of the vertex $v$, and therefore, 
\begin{equation}
\Cl(S)=\Cl_1(S)\cup \{v\} ,
\label{dikla2}
\end{equation}
where $\Cl(\cdot)$ is the closure operation in the original graph $G$ (recall that $G$ contains only one vertex more than
 $G_1$, the vertex $v$).
The size of $S$ satisfies
  $$|S|=|\cup_{i=1}^lS_i|= \sum_{i=1}^l|S_i|\leq r\cdot r^t= r^{t+1}.
  $$
Let us estimate the expansion ratio. By \eqref{eq:Cl} and \eqref{dikla2} 
\begin{align*}
|\Cl(S)|&=|\cup_{i=1}^l\Cl_i(S_i)\cup \{v\}|=1+\sum_{i=1}^l|\Cl_i(S_i)|.
\end{align*}
Hence the expansion ratio of the set $S$ satisfies  
\begin{align}
\frac{|\Cl(S)|}{|S|}&=\frac{1+\sum_{i=1}^l|\Cl_i(S_i)|}{|S|}\nonumber\\
& \geq \frac{1}{r^{t+1}}+ \frac{\sum_{i=1}^l|\Cl_i(S_i)|}{|S|}\nonumber\\
&= \frac{1}{r^{t+1}}+ \sum_{i=1}^l \frac{|S_i|}{|S|}\frac{|\Cl_i(S_i)|}{|S_i|}\nonumber\\
&\geq \frac{1}{r^{t+1}}+ \sum_{i=1}^l \frac{|S_i|}{|S|}e_t \label{dikla3}\\
&=\frac{1}{r^{t+1}}+ e_t\nonumber\\
&=e_{t+1},
\end{align}	 
where \eqref{dikla3} follows since each set $S_i$ has expansion ratio at least $e_t$ in $G_i$. 
\end{IEEEproof}
\begin{lemma}
\label{best lemma2}
Let $m$ be an integer whose base-$r$ representation is
   $$
    m=\sum_{i}\alpha_ir^i,
    $$
then for an integer $t$,
   $$\Big\lfloor \frac{m}{r^t}\Big\rfloor r^t e_t+\sum_{i=0}^{t-1}\alpha_i r^i e_i=\sum_{i=0}^t \Big\lfloor \frac{m}{r^i}\Big\rfloor,
      $$
where the quantity $e_i$ is defined in \eqref{dikla6}.
\end{lemma}
\begin{IEEEproof} 
Note that if $m=\sum_{j\geq0}\alpha_jr^j$ is the r-ary representation of $m$ then
$$r^i \Big\lfloor m/r^i\Big\rfloor=\sum_{j\geq i} \alpha_jr^j,$$ and recall that $$e_t=\frac{r^{t+1}-1}{r^{t+1}-r^t}=\sum_{i=0}^t r^{-i}.$$ 
Then we have that
\begin{align*}
\sum_{i=0}^t \lfloor m/r^i\rfloor&=\sum_{i=0}^t\sum_{j\geq i}\alpha_jr^{j-i}\\
&=\sum_{j\geq 0}\alpha_jr^j \sum_{i=0}^{\min(t,j)}r^{-i}\\
&=\sum_{j= 0}^{t-1}\alpha_jr^je_j +\sum_{j\geq t}\alpha_jr^je_t\\
&=\sum_{j= 0}^{t-1}\alpha_jr^je_j + e_t r^t\lfloor m/r^t\rfloor,
\end{align*}
and the result follows.
\end{IEEEproof}

{\em Remark:} The coloring process of the vertices of $G$ used to construct the closure of the subset is an instance of a large
class of models of influence propagation in networks. Similar models were studied in the literature in a number of contexts related
to random and deterministic graphs. We point to a recent paper \cite{coja15} which studies the minimum size of the subset $S$ of
vertices of a regular expander whose closure under a threshold decision rule equals the entire set of vertices $V$. This paper also contains
pointers to the literature on related problems.

{\vspace*{.1in}{\em Proof of the upper bound on the distance \eqref{eq:d2}:} }
Let $G$ be the recovery graph of the code. 
We will use Lemma \ref{best lemma}  for the graph $G$ several times. 
Assume that we are allowed to color $k-1$ vertices and would like to color them in a way that 
guarantees a large expansion ratio with respect to their closure.
Let $m\leq t$ be the largest integer such that $r^m\leq k-1$, then according to Lemma \ref{best lemma} the graph $H_1:=G$ contains
a subset $S_1$ of vertices of size at most $r^m$ whose 
expansion ratio is at least $e_m$. Color the vertices in $\Cl(S_1).$ 
Then denote by $H_2$ the subgraph induced on the subset of vertices $V\backslash \Cl(S_1)$ and 
apply Lemma \ref{best lemma} to $H_2,$ etc. Continuing this process, suppose that in the $i$-th 
round there are $b_i$ vertices still to be colored (out of a total budget of $k-1$ vertices), and let $H_i$ be the induced subgraph of $G$ 
on the set of vertices that have not been colored in the previous $i-1$ rounds.  
Each vertex in $H_i$ has outgoing edges of all $t$ colors because if not, then one of its recovering
sets has been already removed, but then this vertex itself cannot be present by definition
of the closure. 
Let $m\leq t$ be the largest integer such that $r^m\leq b_i$. 
Now apply Lemma \ref{best lemma} for the graph $H_i$ to find a set $S_i$ of vertices of size at most 
\begin{equation}
|S_i|\leq r^m
\label{dikla4}
\end{equation} 
and expansion ratio at least $e_m$ and color it. Notice that the expansion ratio $e_m$ is an increasing function of $m$, hence in order to get large expansion we would like to choose the largest possible set $S_i$ under the budget constraint.  
Continue this procedure until we have used all the $k-1$ vertices, and call the obtained set of $k-1$
vertices $S.$ Let  $$k-1=\sum_{i}\alpha_ir^i,$$
be the $r$-ary representation of $k-1$. By \eqref{dikla4} the sets $S_i$ in the first $\lfloor \frac{k-1}{r^t} \rfloor$ steps 
of the procedure have expansion ratio at least $e_t$, while the remaining set sets $S_i$ could have a smaller expansion. 
Hence the expansion of the set $S=\cup_i S_i$ is at least  
\begin{align}
\label{qwqw}
|\Cl(S)|\geq \Big\lfloor \frac{k-1}{r^t} \Big\rfloor r^t e_t + \sum_{i=0}^{t-1}\alpha_ir^ie_i.
\end{align}
Using Lemma \ref{best lemma2}, write \eqref{qwqw} as
$$|\Cl(S)|\geq \sum_{i=0}^t\Big\lfloor\frac{k-1}{r^i}\Big\rfloor.$$
By construction, $|S|=k-1$ and so $|\cC_S|\le q^{k-1}$. Clearly also $|\cC_{\Cl(S)}|\le q^{k-1}$, therefore
to finish the proof of \eqref{eq:d2} we take $I=\Cl(S)$ in \eqref{eq:dddd}. 

\section{Lower GV-type bounds for LRC codes}
Here we prove lower asymptotic bounds on the parameters of LRC codes with one and two recovering sets.
The bounds are obtained by studying ensembles of random linear codes with locality properties and rely on a variation of Gallager's method, previously employed for LDPC codes \cite{gal63} and later for bipartite-graph codes \cite{bar06c}.

\subsection{One recovering set, any alphabet}
In this section we prove the lower asymptotic bound on $R_q(r,\delta)$ stated above in Theorem B, Eq.~\eqref{eq0:gv}.

Let $\cC$ be a linear $(n,k,r)$ LRC code over $\ff_q.$
We will use the fact that every code symbol is involved in at least one low-weight parity check of at most $r$ symbols. Suppose that
$n$ is divisible by $r+1.$
Consider an $(n-k)\times n$ matrix over $\ff_q$ of the form $H=\begin{bmatrix}H_U\\H_{L}\end{bmatrix}$ where $H_U$ is a block-diagonal
matrix and $H_L$ has no special structure. Assume that $H_U$ has the form
 \begin{equation}\label{eq:matrix}
 H_U=\begin{bmatrix}
 \text{\framebox[.4in][c]{$H_0$}}&&&\\
 &\text{\framebox[.4in][c]{$H_0$}}&&\\
 &&\ddots&\\
 &&&\text{\framebox[.4in][c]{$H_0$}}
 \end{bmatrix}
 \end{equation}
 where $H_0$ is the parity-check matrix of an $[r+1,r]$ single parity check code (i.e., a row of $r+1$ ones), and the blank spots are filled with 
zeros.  Construct an ensemble of matrices $\cH_q(n,k,r)=\{H\}$ by choosing the elements of $H_L$ uniformly and independently at random from $\ff_q.$ 

We will need the expression for the weight enumerator $b(s)$ of the code with the parity-check matrix $H_0.$
Recall that the weight enumerator of a code of length $n$ is defined as the polynomial 
   \begin{equation}\label{eq:wn}
   A(s)=\sum_{w=0}^n A_w s^w.
   \end{equation}
where $A_w$ is the number of codewords of Hamming weight $w.$ The code with generator matrix $H_0$ contains $q-1$ collinear vectors of weight $r+1$ and the zero vector, and therefore has the weight enumerator $1+(q-1)s^{r+1}.$ 
Using the MacWilliams theorem \cite[p.~146]{mac91} we obtain
    \begin{equation}\label{eq:fg}
    b(s)  = \frac{1}{q}( ( 1 + (q-1)s)^{r+1} + (q-1) (1-s)^{r+1}).
    \end{equation}

\remove{
\begin{itemize}
\item The matrix $H_U$ is a block diagonal, where $H_0$ is a simple parity check code, i.e. a row of $r+1$ ones. 
	\item The entries of $H_L$ are picked randomly and independently from $\mathbb{F}$. Note that we don't need any permutation of the columns.
	\item Using the union bound we show that there exists a matrix $H$ that does not contain in its null space a vector of weight less than $\delta n$.
	\item Using the derived result we bound the number of vectors in the null space of $H_L$ of weight at most $\delta n$ by 
$$|\{x\in \mathbb{F}^n: H_Ux^T=0, \wt(x)<\delta n\}|\leq \delta n \min_{0<s\leq 1}\frac{b(s)^{\frac{n}{r+1}}}{s^{\delta n}}.$$
	\item The probability of nonzero vector to be in the null space of $H_L$ is exactly $q^{-n(\frac{r}{r+1}-R)}$.
	\item Then by the union bound the probability that the code defined by $H$ has distance less than $\delta n$ is at most
	$$\Pr(\exists x: Hx^T=0, \wt(x)<\delta n)\leq \delta n \min_{0<s\leq 1}\{\frac{b(s)^\frac{n}{r+1}}{s^{\delta n}}\}q^{-n(r/(r+1)-R)}$$
\end{itemize}}

\begin{theorem}{\rm (Gilbert-Varshamov bound for LRC codes = Theorem B, Eq.~\eqref{eq0:gv})}\label{thm:gv1}
    \begin{equation}\label{eq:gv}
    R_q(r,\delta)\ge \frac{r}{r+1} - \min\limits_{0<s\le1} \Big\{ \frac{1}{r+1}\log_q b(s) - \delta \log_q s \Big\}.
    \end{equation}
\end{theorem}
\begin{IEEEproof} The code given by the null space of $H_U$ is a direct sum of $n/(r+1)$ single parity check codes, so its weight enumerator equals $b(s)^{n/(r+1)}.$ Let $B_w:=|\{x\in \ff_q^n: \wt(x)=w, H_U x^T=0\}|,$
then it is clear that  
  \begin{equation}\label{eq:aw}
  B_w\le \min_{0<s\leq 1} \frac{b(s)^\frac{n}{r+1}}{s^w}.
  \end{equation}
Note that the value $s=1$ corresponds to the trivial estimate $B_w\le \sum_{w}B_w=q^{\frac{nr}{r+1}}.$ See also the remark after this proof in regards to the
optimization region of $s.$ 
\remove{\textcolor{red}{Furthermore since $0<s\leq 1$ for any  $w_1\leq w_2$ 
\begin{equation}
B_{w_1}\le \min_{0<s\leq 1} \frac{b(s)^\frac{n}{r+1}}{s^{w_1}}\le \min_{0<s\leq 1} \frac{b(s)^\frac{n}{r+1}}{s^{w_2}}.
\label{eq:ten}
\end{equation}}}
    
Let us turn to the matrix $H_L.$ The number of rows of $H_L$ equals $n-k-n/(r+1)= n(r/(r+1)-R),$ and thus
for any nonzero vector $x\in \mathbb{F}_q^n$
    \begin{equation}\label{eq:pl}
    \Pr(H_Lx^T=0)= q^{-n(r/(r+1)-R)}.
    \end{equation}
Using the union bound 
and the fact that the right-hand side of \eqref{eq:aw} grows on $w$ for $0<s\le 1,$
we obtain the estimate
\begin{multline}
  \Pr(\{\exists x\in \mathbb{F}_q^n :Hx^T=0, 0<\wt(x)<\delta n\})\\ 
  \le \delta n q^{-n(\frac r{r+1}-R)} \min_{0<s\le 1}\frac{b(s)^{\frac{n}{r+1}}}{s^{\delta n}}.\label{eq:sum}
\end{multline}
 Thus to prove that the ensemble contains codes with distance $\ge \delta n$ it suffices to show that the right-hand side
of \eqref{eq:sum} is strictly less than one. 
  
Let us compute the logarithm on the right-hand side of \eqref{eq:sum}. We obtain
    \begin{multline}\label{eq:negative}
    \log_q\Big(\delta n q^{-n(\frac r{r+1}-R)} \min_{0<s\le 1}\frac{b(s)^{\frac{n}{r+1}}}{s^{\delta n}}\Big)=
    n\Big(-\frac r{r+1}\\+R+\min_{0<s\le 1}\Big\{\frac 1{r+1}\log_q b(s)
    -\delta\log_q s\Big\}+o(1)\Big).
    \end{multline}
 Choosing $R$ such that for sufficiently large $n$ this quantity becomes negative ensures
 that the probability $\Pr(\{\exists x\in \mathbb{F}_q^n: Hx^T=0, 0<\wt(x)<\delta n\})\to 0$ as $n\to\infty.$   
\end{IEEEproof}

\vspace*{.1in}
{\em Remarks:} 1. It may seem that we are unnecessarily restricting the optimization region in 
the proof 
to $s\in(0,1]$ and that 
by allowing all $s>0$ we may be able to tighten the resulting bound. In the next lemma (proved in the Appendix) we show that this is
not the case, and no loss ensues from this restriction. 

\vspace*{.1in}\begin{lemma}\label{lemma:minimum} Let $b(s)$ be the function defined in \eqref{eq:fg}, and let $1\le d,r\le n$ be positive integers. For $s>0$ there is a unique minimum $\min_s b(s)^{n/(r+1)} s^{-d}$ that is attained for $0<s\le 1.$
\end{lemma}

\vspace*{.1in}
2. From the above proof it is possible to obtain a finite-length lower bound on LRC codes. The following proposition follows from \eqref{eq:sum}.

\vspace*{.1in}\begin{proposition} Let $n,k,r$ be positive integers such that $(r+1)|n$ and $r<k<rn/(r+1).$ If a positive integer $d<n$ satisfies the inequality
    $$d q^{-n\frac r{r+1}+k} \min_{0<s\le 1}\frac{b(s)^{\frac{n}{r+1}}}{s^{d}}<1,$$
there exists a $q$-ary $(n,k,r)$ linear LRC code with distance $d.$ 
\end{proposition}

\vspace*{.1in}This bound is generally better than the direct adaptation of the GV argument given in \eqref{eq:dgv}.

\vspace*{.1in}
\begin{corollary}\label{cor:Plotkin} ({= Theorem B}, Eq. \eqref{eq:pb1}) For any fixed $r,$ $R_q(r,0)=r/(r+1)$ and
    $R_q(r,\delta)=0$ if and only if $\delta\ge (q-1)/q.$
    \end{corollary}
\begin{IEEEproof} From \eqref{eq:Singleton} we obtain that $R_q(r,0)\le r/(r+1)$ while Proposition
\ref{prop:0} implies the reverse inequality. This proves the first statement.
\remove{At the same time, from  \eqref{eq:wn} we have
$b(s)\ge 1$ with equality iff $s=0.$ Using this together with $\delta=0$ in \eqref{eq:gv},  we find that $R_q(r,0)\ge r/(r+1).$}

Next we claim that $R_q(r,\delta)>0$ for all $0\le \delta< (q-1)/q.$ This follows because 
the right-hand side of \eqref{eq:gv} is positive for all $\delta<(q-1)/q.$ The last claim follows from the proof of 
Lemma \ref{lemma:minimum} in the Appendix; see in particular the remarks after the end of the proof.

At the same time, the Plotkin bound (even without the locality constraint)  implies that 
$  R_q(r,\delta)=0$ for any $\delta\ge (q-1)/q.$
This completes the proof.
\end{IEEEproof}

\subsection{Two recovering sets}
Consider LRC codes with two disjoint recovering sets for every symbol. Construct an ensemble of parity-check matrices as
follows. Assume that $n$ is a multiple of $\binom {r+2}2,$ namely, $n=m(r+1)(r+2)/2.$ Let $H$ be the matrix of the form 
$H=\begin{bmatrix}H_U\\H_{L}\end{bmatrix},$ where the submatrices $H_U$ and $H_L$ are of dimensions $m(r+1)\times n$ and $(n-k-m(r+1))\times n,$ respectively. The matrix $H_U$ in \eqref{eq:matrix} is again block-diagonal, but this time the matrix $H_0$ is of dimensions
$(r+1)\times \binom {r+2}2$ and is the edge-vertex incidence matrix of a complete graph $K_{r+2}$ with one row
deleted (deleting the row ensures that the remaining rows are linearly independent). 
The elements of the bottom matrix $H_L$ are chosen uniformly and independently at random from the field $\ff_q.$
The number of rows of the matrix $H_L$ now equals
   $$
   (1-R)n-m(r+1)=\Big(\frac r{r+2}-R\Big)n.
   $$
\begin{lemma} The code with the parity-check matrix $H_0$ has dimension $(r+1)(\frac{r+2}2-1)$ and the weight enumerator 
   \begin{multline}\label{eq:local}
   g_q^{(2)}(s)=\frac1{q^{r+2}}\sum_{\substack {i_0,i_1,\dots,i_{q-1}\\ \sum_{j}i_j=r+2}}
   \binom{r+2}{i_0,i_1,\dots,i_{q-1}}\\ \times(1+(q-1)s)^{\binom{r+2}2-E}(1-s)^E,
   \end{multline}
where 
    \begin{multline}\label{eq:E}
    E= E(i_0,i_1,\dots,i_{q-1})\\\triangleq\begin{cases} {\binom{r+2}2-\sum_{j=0}^{q-1}\binom{i_j}2,} &\text{$q$ even}\\[.1in]
    \binom{r+2}2-\binom{i_0}{ 2}-\frac12\sum_{j=1}^{q-1} i_j i_{q-j}, &\text{$q$ odd. }
    \end{cases}
    \end{multline}
In particular, $g_2^{(2)}(s)$ is given in Eq.~\eqref{eq0:local}.       
\end{lemma}
\begin{IEEEproof}
Even though the formula for $g_2^{(2)}(s)$ can be obtained from \eqref{eq:local}-\eqref{eq:E}, we give an independent proof. On the one
hand, the case $q=2$ is arguably the most interesting; on the other, it makes it easier to understand the general argument. 
Below
by $\tilde H_0$ we denote the full edge-vertex incidence matrix of the graph $K_{r+2},$ before one row is deleted from it to
obtain $H_0$. Its dimensions are
 $(r+2)\times\binom{r+2}2$ and $\rank(\tilde H_0)=r+1.$
    
\vspace*{.05in}    {\em Case $q=2$}. Consider the code $\cC^\bot$ spanned by the rows of $\tilde H_0$ over $\ff_2.$ Let $V_1$ be a subset of vertices of $K_{r+2}$ and consider the codeword $x$ given by a sum of rows that correspond to the vertices in $V_1.$ Each coordinate corresponds to an edge, and any edge with both ends in $V_1$ or both ends in $V_1^c$ accounts for a zero coordinate of $x$.
Moreover, the nonzero coordinates
are precisely those edges with one end in $V_1$ and the other in $V_1^c.$ Thus, the weight enumerator of the
code $\cC^\bot$ equals
    $$
     g^\bot(y)=\frac12 \sum_{i=0}^{r+2}\binom{r+2}i y^{i(r+2-i)}.
    $$
    The factor $1/2$ accounts for the fact that $\rank (\tilde H_0)=r+1$,  so the above procedure
counts every code vector twice, once for the subset $V_1$ and the second time for $V_1^c.$
    
 The expression for $g_2^{(2)}(s)$ now follows on applying the MacWilliams theorem \cite[p.~146]{mac91}.   
 
\vspace*{.05in}  {\em Arbitrary $q$.} Let $\ff_q=\{\alpha_0=0,\alpha_1,\dots,\alpha_{q-1}\},$ where for $q$ odd the numbering of the elements of
 the field is such that $\alpha_i=-\alpha_{q-i}.$
 
Consider 
 the code $\cC^\bot$ spanned by the rows of $\tilde H_0$ (or of $H_0$) over $\ff_q.$ 
Let $x$ be a codeword in $\cC^\bot$
 and assume that $x$ is a linear combination of the rows of $\tilde H_0$ with coefficients $a_1, a_2,\dots,a_{r+2}.$
The coordinates of $x$ correspond to the edges, and $x_j=0$ if and only if the ends of the $j$th edge add to zero, i.e., if
$a_{j_1}+a_{j_2}=0,$ where $j_1,j_2$ and the vertices connected by edge $j.$ Let
   $$
   i_j:=|\{l\in\{1,2,\dots,r+2\}: a_l=\alpha_j\}|, \quad j=0,1,\dots,q-1
   $$ 
be the composition of the coefficient vector.   
If $q$ is even, then the coordinate $x_j=0$ if and only if $a_{j_1}=a_{j_2}.$   If $q$ is odd, then $x_j=0$ if and only if either
$a_{j_1}=a_{j_2}=0$ or $0\ne a_{j_1}=-a_{j_2}.$
Suppose that the composition $(i_0,i_1,\dots,i_{q-1})$ is fixed. Then for even $q$ the number of nonzero coordinates $x_j$ is given by
the first of the two expressions for $E(i_0,i_1,\dots,i_{q-1}),$ while for odd $q$ it is given by the second expression. 

Recalling that $\rank(\tilde H_0)=r+1,$ we see that every codeword was counted $q$ times. Therefore, the weight enumerator
of the code $\cC^\bot$ equals
     $$
     g^\bot(y)=\frac 1q \sum_{\substack {i_0,i_1,\dots,i_{q-1}\\ \sum_{j}i_j=r+2}}\binom{r+2}{i_0,i_1,\dots,i_{q-1}}
        y^E.
     $$
The proof is finished by the application of the MacWilliams theorem.   
\end{IEEEproof}

\vspace*{.1in} 

The following theorem is proved by computing the expected distance of the code in the ensemble given by the matrices $H$.

\begin{theorem}{\rm (Gilbert-Varshamov bound for $2$-LRC codes = Theorem B, Eqns.\eqref{eq0:gv2} and \eqref{eq:pbm})}
   \begin{equation}\label{eq:gv2}
R_q^{(2)}(r,\delta) \ge \frac{r}{r+2} - \min_{0<s\le 1} 
\Big\{ \frac{1}{\binom{r+2}{2}}\log_q g_q^{(2)}(s) - \delta \log_q s \Big\}. 
   \end{equation}
In particular, $R_q^{(2)}(r,0)\ge\frac r{r+2}$ and $R_q^{(2)}(r,\delta)=0$ if and only if $\delta\ge\frac{q-1}q.$ 
\end{theorem}
\begin{IEEEproof}
First let us show that every coordinate of the null space of $H$ has two disjoint recovering sets of size $r+1.$
Consider a subset of coordinates of size $\binom{r+2}2$ that corresponds to one instance of $H_0.$
Every edge in the graph $K_{r+2}$ is connected to two vertices, and the rows of $H_0$ that contain these
vertices, contain two sets of ones that intersect only on the chosen edge. These sets form the recovering
sets of the chosen edge, and they are obviously disjoint (because the graph does not contain multiple edges).

The remaining part of the proof is computational. It follows the steps \eqref{eq:wn}-\eqref{eq:negative} and is completely analogous to the proof of Theorem \ref{thm:gv1} and Corollary \ref{cor:Plotkin}.\end{IEEEproof}

\remove{\begin{figure}[ht] \vspace*{.2in}\centerline{\includegraphics[scale=0.5]{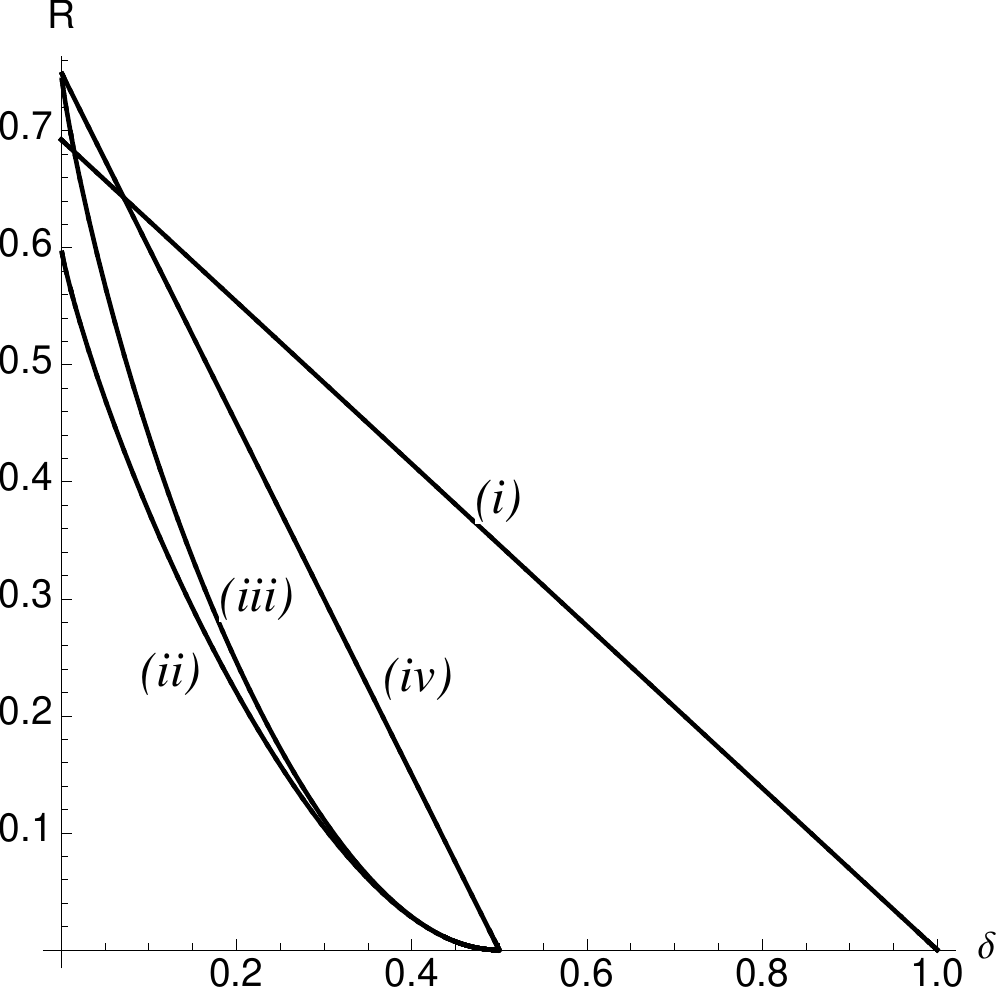}} 
\caption{The GV bounds for LRC codes, $q=2$, $r=3$ for one and two recovering sets. Note that $R_q(r,\frac{q-1}q)=R^{(2)}_q(r,\frac{q-1}q)=0$. The top curve is the GV bound with no locality constraints.} \label{fig2}
  \end{figure}
}

\subsection{Multiple recovering sets, Large alphabets. Proof of Theorem C}

In this section we show the existence of an $(n,k,r,t)$ LRC codes over a sufficiently
large finite field $\ff_q$ with large minimum distance  and rate 
\begin{equation}
R\leq 1- \frac{t}{r+1}.
\label{eq:99}
\end{equation} 
The proof relies on the existence of regular bipartite graphs with good expansion properties. 

For a subset $A\subseteq [m]$ 
construct a vector $x_A=(x_{1,A},...,x_{m,A})\in \ff_q^m$ as follows: 
  \begin{equation}\label{eq:Rc}
  x_{i,A}=\begin{cases} 0 &\text{if }i\notin A\\ X_{i,A} &\text{if }i\in A\end{cases}
  \end{equation}
where $X_{i,A}$ are independent random variables uniformly distributed over $\ff_q^\ast\triangleq \ff_q\backslash\{0\}.$ 

Recall that a family of subsets of an $m$-set $\cS\subset 2^{[m]},\cS=\{A_1,...,A_{|\cS|}\}$  is said to satisfy {\em Hall's condition}\
if for any subfamily $\cS'\subseteq \cS $, 
    \begin{equation}\label{eq:H}
    |\cS'|\leq |\cup_{A\in \cS'} A|.
    \end{equation}
By Hall's matching theorem, such a family contains a system of distinct representatives, i.e., a set of $|\cS|$ distinct elements  $a_i\in [m]$ such that $a_i\in A_i$ for each $i$.     
\vspace*{.1in}
\begin{lemma} \label{lemma:Hall} Let $\cS$ be a family of subsets of an $m$-set that satisfies Hall's condition. For a sufficiently large
$q$ the vectors $\{x_{A}, A\in \cS\}$ of the set defined by $\cS$ are linearly independent with positive probability which tends to one
as $q\to \infty.$
\end{lemma}
\vspace*{.1in}
\begin{IEEEproof} Let $|\cS|=s$ and consider the $m\times s$ matrix $M$ whose columns are the vectors $\{x_{A}, A\in \cS\}$. Note that 
\eqref{eq:H} implies that $s\leq |\cup_{A\in S} A|\leq m.$ We will show that with high probability the rank of the matrix $M$ is $s$. Since $
\cS $ contains a system of distinct representatives, there exists an injective mapping $f$ from the columns of $M$ to its set of rows such that the random variable 
$X_{f(A),A}$ is not identically zero for any $A\in \cS$. Let $M'$ be the $s\times s$ submatrix of $M$ restricted to the rows $f(A),A\in \cS.$ 
The determinant of this matrix is a homogeneous polynomial of total degree $s$ that is not identically zero: for instance, it contains the 
nonzero term $\prod_{A\in \cS }x_{f(A),A}$. By the Schwartz-Zippel Lemma \cite{schwartz80} we have that $P(\det M'=0)\le s/|\ff_q^\ast|,$ where
the probability is computed with respect to the random choice in \eqref{eq:Rc}.
In other words, 
   $$
   P(\rank M=s)\ge P(\rank(M')=s)=1-\frac{s}{q-1}.
   $$
\end{IEEEproof}

Let $G=(V=V_1\cup V_2, E)$ be a biregular bipartite graph with $\deg v=t$ for $v\in V_1$ and $\deg v=r+1$ for $v\in V_2.$
Suppose that $|V_1|=n,$ then the number of vertices in $V_2$ is $p\triangleq nt/(r+1).$
 The graph $G$ is called an $(t,r+1,\delta, \gamma)$-\emph{expander} if every subset $T\subset V_1, |T|\le \delta n$ 
has at least $\gamma t|T|$ neighbors in $V_2,$ where $0< \delta,\gamma\leq 1.$ 
The following result, due to \cite{bur01}, is cited here in the form given in \cite[p.~431]{RU2008}.
\vspace*{.1in}
\begin{lemma} %
\label{bursh}
Let $G$ be a graph chosen uniformly from the ensemble of $(t,r+1)$-regular bipartite graphs and let $n\to\infty.$ For a given $\gamma\in [\frac 1{r+1},1-\frac{1}{t})$ let $\delta$ 
be the positive solution of the equation 
  \begin{equation}
 \frac{t-1}{t} h(\delta)-\frac{1}{r+1} h(\delta\gamma(r+1))-\delta\gamma(r+1) h\Big(\frac{1}{\gamma(r+1)}\Big)=0.
\label{eq:1.1}
 \end{equation} 
Then for $0<\delta'<\delta$ and $\beta=t(1-\gamma)-1$
\begin{equation}
\Pr(\{G \text{ is an} (t,r+1,\delta',\gamma) \text{ expander}\})\geq 1-O(n^{-\beta}).
\label{eq:bursh1}
\end{equation}
\end{lemma}
Note that the conditions on $\gamma$ in \cite{RU2008} are stated as $0\le\gamma<1-1/t,$ but the last entropy function in \eqref{eq:1.1} is not 
defined for $\gamma<1/(r+1).$ {However, for such $\gamma$ any subset of vertices $T\subseteq V_1$ has at least $t|T|/(r+1)\geq \gamma t|T|$ distinct neighbors in $V_2$ and therefore any  $(t,r+1)$-regular bipartite graph is an $(t,r+1,1, \gamma)$-\emph{expander}.} 

Given a bipartite biregular graph $G$, define a family of subsets $\cS$ of the set $[n-k]$ as follows. 
Assume that the vertices in $V_1$ (in $V_2$) are numbered from $1$ to $n$ (from $1$ to $p$). 
For every vertex $i\in V_1$ form a subset $S_i=N_i\cup [p+1,n-k]$, where $N_i\subset V_2$ is the set of neighbors
of $i$ in $V_2.$ Note that on account of \eqref{eq:99} each $S_i$ is indeed a subset of $[n-k]$.

\vspace*{.1in}
\begin{lemma}
\label{existence}
Let $G$ be an $(t,r+1,\delta, \gamma)$ expander, where the variables $\delta, \gamma$ are the \emph{unique} solution of   
\eqref{eq:1.1} and the equation
\begin{equation}
\delta(1-t\gamma)={1-\frac{t}{r+1}-R},
\label{eq:0.1}
\end{equation} in the range    
$  \gamma\in [\frac 1{r+1},\frac 1t),
    $  
where $R\le 1-t/(r+1).$
Consider the sets $S_i$ defined above (before the statement of the lemma).
Then any family of subsets $\cS =\{S_i:i\in I\}$ of size $|\cS|\le \delta n$
satisfies Hall's condition.
\end{lemma}

\begin{IEEEproof} The proof is formed of two steps. First we show that the 
$(t,r+1,\delta, \gamma)$ expander graph $G$ is well defined,
i.e., that there is a pair of numbers $(\delta,\gamma)$ that satisfies \eqref{eq:1.1}
and \eqref{eq:0.1}. Each of these equations defines $\delta$ as a continuous function of $\gamma.$ 
Let $\delta_1(\gamma)$ be the function defined by \eqref{eq:0.1} and let $\delta_2(\gamma)$ be defined by \eqref{eq:1.1} (as argued in \cite{RU2008}, $\delta_2$ is well defined in the sense that Eq.~\eqref{eq:1.1} has a unique positive root $\delta$). The function $\delta_1(\gamma)$ increases monotonically from a number less than 1 to $+\infty$ as $\gamma$ ranges from $\frac 1{r+1}$ to $1/t.$ At the same time, the value $\delta_2(1/(r+1))$ is determined by the equation
    $$
    h(\delta)\Big(\frac{t-1}{t} -\frac{1}{r+1}\Big)=0.
    $$
Since $t\geq 2$ and $\frac{t}{r+1}<1,$ we conclude that $h(\delta)=0$, and since $\delta\ne 0,$ this implies that $\delta_2(1/(r+1))=1.$
From Lemma \ref{bursh} and since $t\geq 2$, for any $\gamma\in  [\frac 1{r+1},\frac{1}{t})$ there exists an expander graph for all $\delta'<\delta_2(\gamma).$ This 
implies that $\delta_2(\gamma)$ is a bounded function for $\gamma$ in this range; {in fact, it is easy to check that $\delta_2(\gamma)$ is a monotonically decreasing function}. Therefore, there exists \emph{exactly one} one
$\gamma\in [\frac 1{r+1},\frac 1t)$ such that $\delta_1(\gamma)=\delta_2(\gamma).$

Let us prove the claim about Hall's condition. Let $I\subseteq [n]$ be a subset of indices of size $\delta' n < \delta n$.
\begin{align*}
|\cup_{i\in I}S_i|&=|\cup_{i\in I}N_i\cup [p+1,n-k]|\\
&=|\cup_{i\in I}N_i|+n-k-p\\
&\geq t\gamma \delta' n+n-k-p\\
&=n(t\gamma \delta'+1-R-\frac{t}{r+1})\\
&\geq \delta' n,
\end{align*}
where the first inequality follows from the expansion property of $G$ and the second inequality follows from 
\eqref{eq:0.1} and the fact that $\delta' \leq \delta$.
\end{IEEEproof}

Now we are ready to complete our argument.
\begin{IEEEproof} (of Theorem C)
Let $G$ be the graph in Lemma \ref{existence} and let the corresponding  family of subsets be $\cS =\{S_1,...,S_n\}$. 
On account of Lemma \ref{lemma:Hall}, there exists a set of $n$ vectors of length $n-k$ such that any $\delta n$ of them are linearly independent. Let these vectors form the parity check matrix $H$ of a code $\cC$. 
Then it is clear that the minimum distance of that code is at least  $\delta n$. Moreover, the first $p$ rows of the matrix 
$H$ are of weight $r+1$ and they provide the locality property for the code's symbols. 

Observe that the recovering sets defined by this construction are not necessarily disjoint but become such if the graph contains
no cycles of length 4. As shown in \cite{mckay04}, the probability that a random regular graph on $n$ vertices has no cycles of length 4 is \emph{bounded away} from zero as $n\to\infty.$ As argued in the last section \cite{mckay04}, the methods of that paper apply to bipartite
graphs, leading to a similar conclusion. At the same time, Lemma \ref{bursh} implies that the probability for a random graph to have the claimed
expanding properties approaches one. Together these results imply that there exist $(t,r+1,\delta, \gamma)$ biregular bipartite expanding graphs with
no cycles of length 4, i.e., that there exist $(n,k,r,t)$ LRC codes with the stated parameters. This concludes the proof.
\end{IEEEproof}

\vspace*{.1in}
\section{Concluding remarks}
The problem of bounding the cardinality of LRC codes with a given distance poses a number of interesting challenges even for a single
recovering set. While the asymptotic version of this problem is presently in the same state as the asymptotic problem of bounding the
size of error correcting codes without the locality constraint, for finite parameters the only meaningful bound that accounts for the
size of the alphabet is the shortening bound of \cite{Cad13a}. We believe that the asymptotic GV-type bound is in a certain
sense ``in a final form,'' i.e., this bound cannot be improved by studying ensembles of random codes without bringing in significant new
ideas. At the same time, the field for the upper bounds seems to be open in the sense that it should be possible to find bounds that improve on the currently known results. In particular, since the structure of LRC codes shows some similarity to LDPC codes, 
it is likely that methods of deriving upper bounds on LDPC codes could yield good upper bounds on LRC codes. We note that straightforward application of techniques developed for LDPC codes, e.g., \cite{bhl06}, does not lead to improved upper bounds in the LRC case.

For codes with multiple recovering sets it is difficult to derive good lower or upper bounds because there is little control over the
structure of the sets. Nevertheless, we believe that the GV-type bound for $t=2$ derived in this paper will be difficult to improve by
studying ensembles of random codes. At the same time, it could be possible to use constructions on algebraic curves to obtain improvements
of the GV bound for $t\ge 2$. For the case $t=1$ such improvements were obtained in the recent work \cite{barg15a}. 

Finally, and interesting open question is to establish (or disprove) the tightness of the Singleton-like bound \eqref{eq:sa} for
multiple recovering sets in the case of large alphabets. A related research direction is derandomizing the expander graph lower
bound derived in this paper.

\appendix
{\sc Proof of Lemma \ref{lemma:minimum}.} Consider the function
   $$
   F(s)=\frac 1{r+1}\ln b(s)-\delta\ln s.
   $$
The lemma will be proved if we show that for every $\delta\in (0,(q-1)/q],$ $F(s)$ has a unique minimum attained
for $0< s\le 1.$ Setting $F'(s)=0,$ we obtain the equation
  \begin{equation}\label{eq:=}
  f(s)=\delta,
  \end{equation}
   where
  $$
  f(s)=\frac 1{r+1}\frac {s b'(s)}{b(s)}.
  $$ 
It is easy to check that \eqref{eq:=} has the solutions $(s=0,\delta=0)$ and $(s=1,\delta=(q-1)/q).$
Suppose we prove that $f(s)$ is strictly monotone increasing for $s\ge 0.$ This would imply that
the inverse function $s=f^{-1}(\delta)$ is also strictly monotone increasing on $\delta,$ and therefore,
the minimizing value $f^{-1}(\delta)$ for all $0<\delta<(q-1)/q$ is unique and is located in the open
interval $(0,1).$

It remains to prove that $f(s)$ is indeed a strictly increasing function of $s\ge 0.$ We have
   $$
      f'(s)=\frac 1{r+1}\frac {(s b'(s))' b(s)-s (b'(s))^2}{b(s)^2}.
   $$   
Recalling that $b(s)$ is a polynomial of $s$ of degree $r+1$, let us write it as
  $$
   b(s)=1+\sum_{i=2}^{r+1} b_is^i.
  $$
Next
  \begin{multline*}
  s(s b'(s))' b(s)-(s b'(s))^2\\=\Big(\sum_{i=2}^{r+1}i^2b_is^i\Big)\Big(1+\sum_{i=2}^{r+1}b_is^i\Big)-\Big(\sum_{i=2}^{r+1}ib_is^i\Big)^2.
  \end{multline*}
The right-hand side of this equality is positive for all $s>0$ if 
  $$
  \Big(\sum_{i=2}^{r+1}i^2b_is^i\Big)\Big(\sum_{i=2}^{r+1}b_is^i\Big)-\Big(\sum_{i=2}^{r+1}ib_is^i\Big)^2\ge0.
  $$
But this last claim is simply the Cauchy-Schwartz inequality for the real vectors $(i\sqrt {b_is^i})_{i=2}^{r+1}$
and $(\sqrt {b_is^i})_{i=2}^{r+1}.$ The lemma is proved.

  Note that since the minimizing value $s_0$ is less than $1$ for all $\delta<(q-1)/q,$ and since
the value of the minimum in \eqref{eq0:gv} increases from 0 to $r/(r+1)$ as $s_0$ ranges
from $0$ to $1,$ the right-hand side of \eqref{eq0:gv} is positive for all $\delta\in(0,(q-1)/q).$

\bibliography{LRC_bib}
\bibliographystyle{IEEEtranS}

\vspace*{.2in}

{\small
{{\bf Itzhak Tamo} was born in Israel in 1981. He received a B.A. in Mathematics and a B.Sc. in Electrical Engineering in 2008, and
a Ph.D. in Electrical Engineering in 2012, all from Ben-Gurion University, Israel. During 2012-2014 he was a postdoctoral researcher
at the Institute for Systems Research, University of Maryland, College Park. Since 2015 he has been a senior lecturer in
the Electrical Engineering Department, Tel Aviv University, Israel.

Itzhak Tamo received the 2015 IEEE Information Theory Society
Paper Award and the IEEE Communication Society Data Storage Technical
Committee 2013 Best Paper Award. His research interests include storage systems and devices, coding, information theory,
and combinatorics.\par}

\vspace*{.2in}

{\small{\bf Alexander Barg} (M'00-SM'01-F'08) received the M.Sc. degree in applied
mathematics and the Ph.D. degree in electrical engineering, the latter from
the Institute for Information Transmission Problems (IPPI) Moscow, Russia,
in 1987. He has been a Senior Researcher at the IPPI since 1988.
 Since 2003 he has been a Professor
in the Department of Electrical and Computer Engineering and Institute for
Systems Research, University of Maryland, College Park.

Alexander Barg was a co-recipient of the IEEE Information Theory Society
Paper Award in 2015. During 1997-2000, A. Barg was an Associate Editor
for Coding Theory of the IEEE Transactions on Information Theory.
He was the Technical Program Co-Chair of the 2006 IEEE International
Symposium on Information Theory and of 2010 and 2015 IEEE ITWs.
He serves on the Editorial Board of several journals including {\em Problems
of Information Transmission}, {\em SIAM Journal on Discrete Mathematics}, and
{\em Advances in Mathematics of Communications}.

Alexander Barg's research interests are in coding and information theory,
signal processing, and algebraic combinatorics.\par}

\vspace*{.2in}

{\small\textbf{Alexey Frolov} was born in Moscow, Russia, in 1987. He received his Specialist Diploma from Bauman Moscow State Technical University (BMSTU) in 2010, and his Ph.D. (Candidate of Science) degree from the Institute for Information Transmission Problems of the Russian Academy of Sciences (IITP RAS) in 2012. Currently, he is a senior researcher at the IITP RAS, Moscow, Russia. His research interests include coding theory and its applications in telecommunications, storage systems and other areas.
\par}

\vfill

\end{document}